\documentclass[journal]{IEEEtran}
\usepackage{colortbl}
\usepackage{ifpdf}
\usepackage{cite}
\usepackage{amsmath,amssymb,amsfonts}
\usepackage{algorithmic}
\usepackage{graphicx, adjustbox}
\usepackage{array}
\usepackage{enumitem}
\usepackage{breqn}
\usepackage{ dsfont }
\usepackage{ upgreek }
\usepackage{textcomp}
\usepackage{multirow}
\usepackage{algorithm}
\usepackage{algorithmic}
\usepackage{caption}
\usepackage{pgfplots}
\usepackage{tkz-euclide,subfigure}
\usepackage{tikz}
\usepackage{balance}
\newcolumntype{P}[1]{>{\centering\arraybackslash}p{#1}}
\begin{document}
	
	\title{Integration of Blockchain and Cloud of Things: Architecture, Applications and Challenges}
	
	\author{Dinh C. Nguyen,~\IEEEmembership{Student Member,~IEEE,}
		Pubudu N. Pathirana,~\IEEEmembership{Senior Member,~IEEE,} \\
		Ming Ding,~\IEEEmembership{Senior Member,~IEEE,} and
		Aruna Seneviratne,~\IEEEmembership{Senior Member,~IEEE}
		
		\thanks {*This work was supported in part by the CSIRO Data61, Australia.}
		\thanks{Dinh C. Nguyen is with School of Engineering, Deakin University, Waurn Ponds, VIC 3216, Australia, and also with the CSIRO Data61, Docklands, Melbourne, Australia  (e-mail: cdnguyen@deakin.edu.au).}% <-this % stops a space
		\thanks{Pubudu N. Pathirana is with School of Engineering, Deakin University, Waurn Ponds, VIC 3216, Australia (email: pubudu.pathirana@deakin.edu.au).}
		\thanks{Ming Ding is with the CSIRO Data61, Australia (email: ming.ding@data61.csiro.au).}% <-this % stops a space
		\thanks{Aruna Seneviratne is with School of Electrical Engineering and Telecommunications, University of New South Wales (UNSW), NSW, Australia (email: a.seneviratne@unsw.edu.au).}
	}
	
	\markboth{IEEE Communications Surveys \& Tutorials}%
	{IEEE Communications Surveys \& Tutorials}
	\maketitle
	
	\begin{abstract}
The blockchain technology is taking the world by storm. Blockchain with its decentralized, transparent and secure nature has emerged as a disruptive technology for the next generation of numerous industrial applications. One of them is Cloud of Things enabled by the combination of cloud computing and Internet of Things. In this context, blockchain provides innovative solutions to address challenges in Cloud of Things in terms of decentralization, data privacy and network security, while Cloud of Things offer elasticity and scalability functionalities to improve the efficiency of blockchain operations. Therefore, a novel paradigm of blockchain and Cloud of Things integration, called BCoT, has been widely regarded as a promising enabler for a wide range of application scenarios. In this paper, we present a state-of-the-art review on the BCoT integration to provide general readers with an overview of the BCoT in various aspects, including background knowledge, motivation, and integrated architecture. Particularly, we also provide an in-depth survey of BCoT applications in different use-case domains such as smart healthcare, smart city, smart transportation and smart industry. {Then, we review the recent BCoT developments with the emerging blockchain and cloud platforms, services, and research projects.} Finally, some important research challenges and future directions are highlighted to spur further research in this promising area. 
	\end{abstract}
	
	
	\IEEEpeerreviewmaketitle

	\section{{Introduction}}
	
	Recent years have witnessed the explosion of interest in blockchain, across a wide span of applications from cryptocurrencies to industries \cite{1}, \cite{3}.  The rapid development in the adoption of blockchain as a disruptive technology is paving the way for the next generation of financial and industrial service sectors. Indeed, new research activities on blockchain and its applications take place every day, impacting many aspects of our lives, such as finance \cite{5}, energy \cite{6}, and government services \cite{7}. 
	
	From a technical perspective, blockchain is a distributed ledger technology that was first used to serve as the public digital ledger of cryptocurrency Bitcoin \cite{8} for economic transactions. The blockchain is basically a decentralized, immutable and public database. The concept of blockchain is based on a peer-to-peer network architecture in which transaction information is not controlled by any single centralized entity. Transactions stored in a chain of blocks are publicly accessible to all blockchain network members in a trustworthy manner. Blockchain uses consensus mechanisms and cryptography to validate the legitimacy of data transactions, which guarantees resistance of linked blocks against modifications and alterations \cite{9}. In particular, the blockchain technology also boasts the desirable characteristics of decentralization, accountability, and security which improve service efficiency and save operational costs. Such exceptional properties promote the usage of applications built on blockchain in recent years. Thus, it makes now the right time to pay attention to this hot research topic.
	\begin{figure*}
		\centering
		\includegraphics[height=5.3cm, width=18cm]{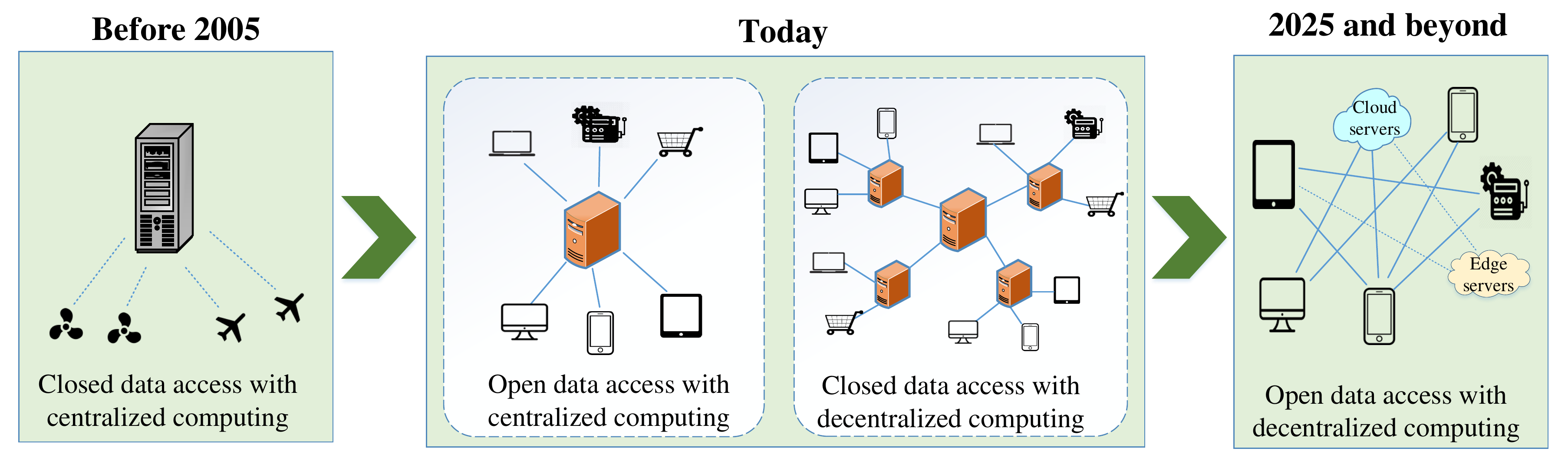}
		\caption{Past, present and future Cloud of Things infrastructure.   }
		\label{CoT1}
		\vspace{-0.17in}
	\end{figure*}
	
	On the other side, the revolution in the field of information and communication has created a wealth of opportunities for advanced technologies, especially Internet of Things (IoT) and Cloud computing.  IoT has reshaped and transformed our lives with various new industrial, consumer, and commercial services and applications \cite{10}, \cite{11}. Typically, IoT is a system of physical objects that can be monitored, controlled or interacted with by ubiquitous electronic devices to enable ubiquitous industrial services, e.g., smart cities, smart industries, etc. Due to the limited resources of IoT devices, they always delegate IoT application tasks to Cloud computing, which gives birth to the Cloud of Things (CoT) paradigm \cite{12}, \cite{13}. The CoT provides a flexible, robust cloud computing environment for processing and managing IoT services, showing great potentials to improve the system performance and efficiency of service delivery  \cite{14}. 
	However, the conventional CoT infrastructures tend to be ineffective due to the following challenges. First, the conventional CoT solutions have mainly relied on centralized communication models, e.g., central cloud, for IoT service operations which make it hard to scale when IoT networks become more widespread \cite{15}. Moreover, most current CoT systems mandate trusting a third party, e.g., a cloud provider, for IoT data processing, which raises data privacy concerns. Final, the centralized network infrastructure results in higher communication latency and power consumption for IoT devices due to long data transmission, which hinders the large-scale deployments of CoT in practical scenarios \cite{16}. 
	
In order to achieve a sustainable development of CoT, building a more decentralized ecosystem has been regarded as a future direction to replace centralized computing models used in current applications, as illustrated in Fig.~\ref{CoT1}. It is strongly believed that blockchain will be a strong candidate to realize the full decentralization of future CoT networks. Particularly, the integration of blockchain and CoT leads to a novel paradigm called as \textit{BCoT}. The combination of these emerging technologies brings great benefits to both worlds and thus gains sustainable interest in academics and industries.
{In fact, the blockchain and CoT have a number of complementary connections for practical applications. In the context of cloud computing, blockchain has been regarded as a service called Blockchain as a Service (BaaS). By providing a decentralized storage architecture using virtual storage nodes, blockchain can enable completely new cloud storage functions which are strongly resistant to data modifications. Instead of relying on traditional cloud data centres, blockchain interconnects computer nodes, including virtual machines on cloud and external computers, to build a fully decentralized storage system without requiring a central authority. Blockchain also functions as network management services which are closely related to smart contract-based applications. In such scenarios, blockchain acts as a communication layer among cloud servers, IoT devices, and end users.} Specifically, the adoption of blockchain can provide many potential benefits for CoT systems as follows.
	\begin{itemize}
		\item Decentralization: Blockchain with its decentralized nature is a promising methodology to effectively solve the bottleneck and single-point failure issues by eliminating the requirement for a trusted third party in the CoT network \cite{17}. Further, the peer-to-peer architecture of blockchain allows all network participants to verify IoT data correctness and ensure immutability with equal validation rights.  
		\item Security and Privacy: The BCoT system can achieve a trustworthy access control by using blockchain-enabled smart contracts \cite{18} which enable to authorize automatically all operations of cloud providers and IoT devices and prevent potential threats to cloud resources and enhance fine-grained control on IoT data \cite{19}. Moreover, the blockchain enables users to track their transactions over the network so as to maintain device and data ownership for improving information privacy.
		\item Corporation: Blockchain enables a new cooperation ecosystem among multiple entities with unlimited data sharing capabilities without trusting each other. The removal of the third party helps establish open environments where any IoT users and cloud providers interested in the system can participate and collaborate to achieve the common goals within the BCoT ecosystem \cite{20}. 
		
	\end{itemize}
	On the other hand, CoT can support blockchain platforms with the following key benefits:
	\begin{itemize}
		\item Scalable support for blockchain transactions: In large-scale blockchain applications, the number of transactions in blockchain networks can be enormous. Therefore, it is highly necessary to provide powerful data processing services to accelerate transaction execution in order to enable scalable blockchain services. In this context, the cloud can offer on-demand computing resources for blockchain operations thanks to its elasticity and scalability capability \cite{21}. For example, public clouds can offer a large-scale network of resources for blockchain service operators in a federated cloud environment. Therefore, the combination of cloud computing and blockchain can achieve a high scalability of the integrated system.
		\item Fault tolerance: Cloud can help replicate blockchain data across a network of computing servers which are interconnected robustly by collaborative clouds \cite{22}. This will minimize the single-failure risks due to the disruption of any cloud node and thus ensure uninterrupted services. Further, the inter-cloud ecosystem can enable the blockchain system to operate continuously in the event of a certain cloud server being under attack. 
	\end{itemize}
	
	\begin{figure}
		\centering
		\includegraphics[ height=6cm, width=9cm]{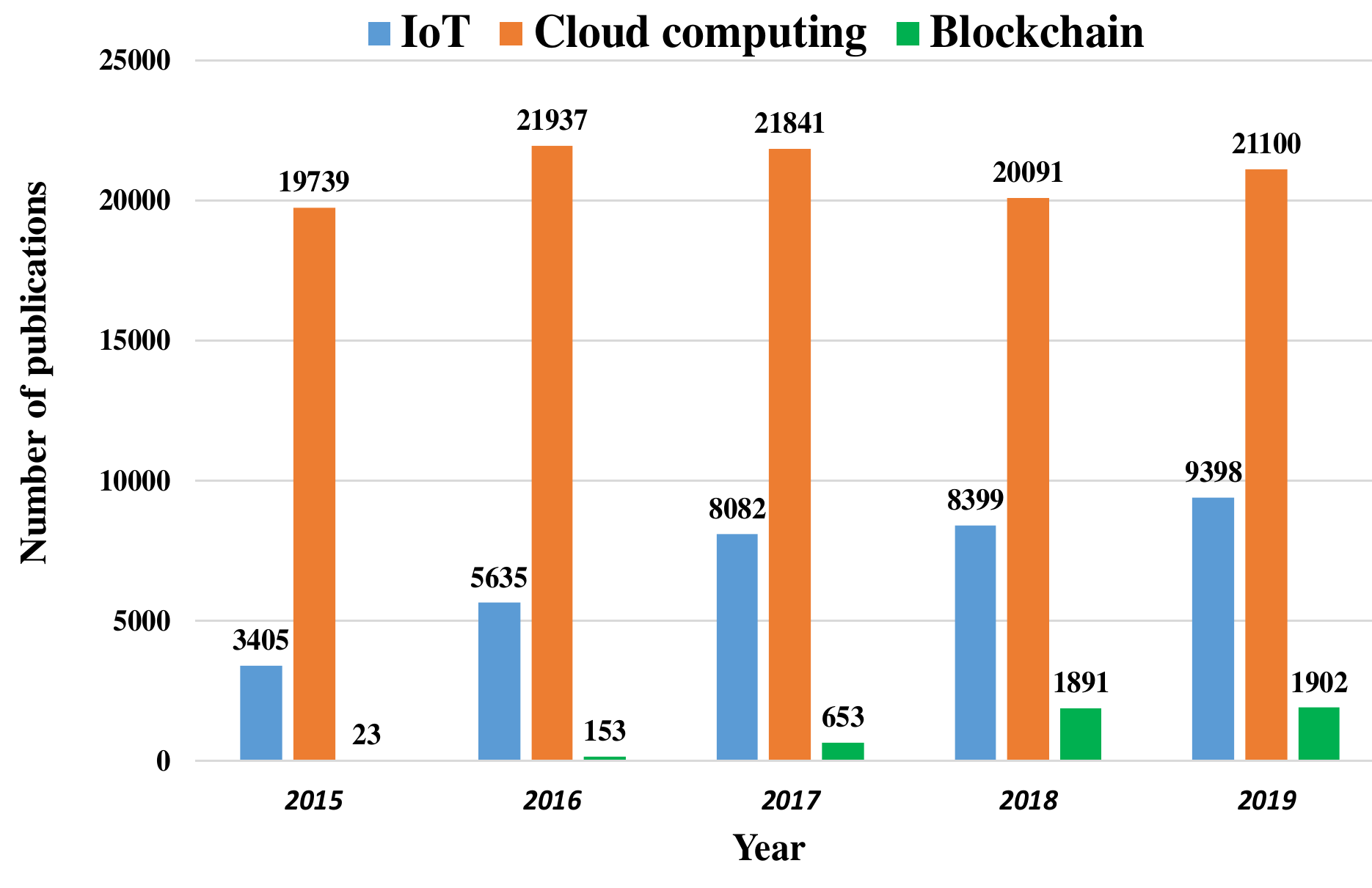}
		\caption{Research trends about IoT, Cloud and blockchain (Source: Web of Science).}
		\label{researchtrend}
		\vspace{-0.15in}
	\end{figure}
\begin{figure*}
	\centering
	\includegraphics[height=13cm, width=13cm]{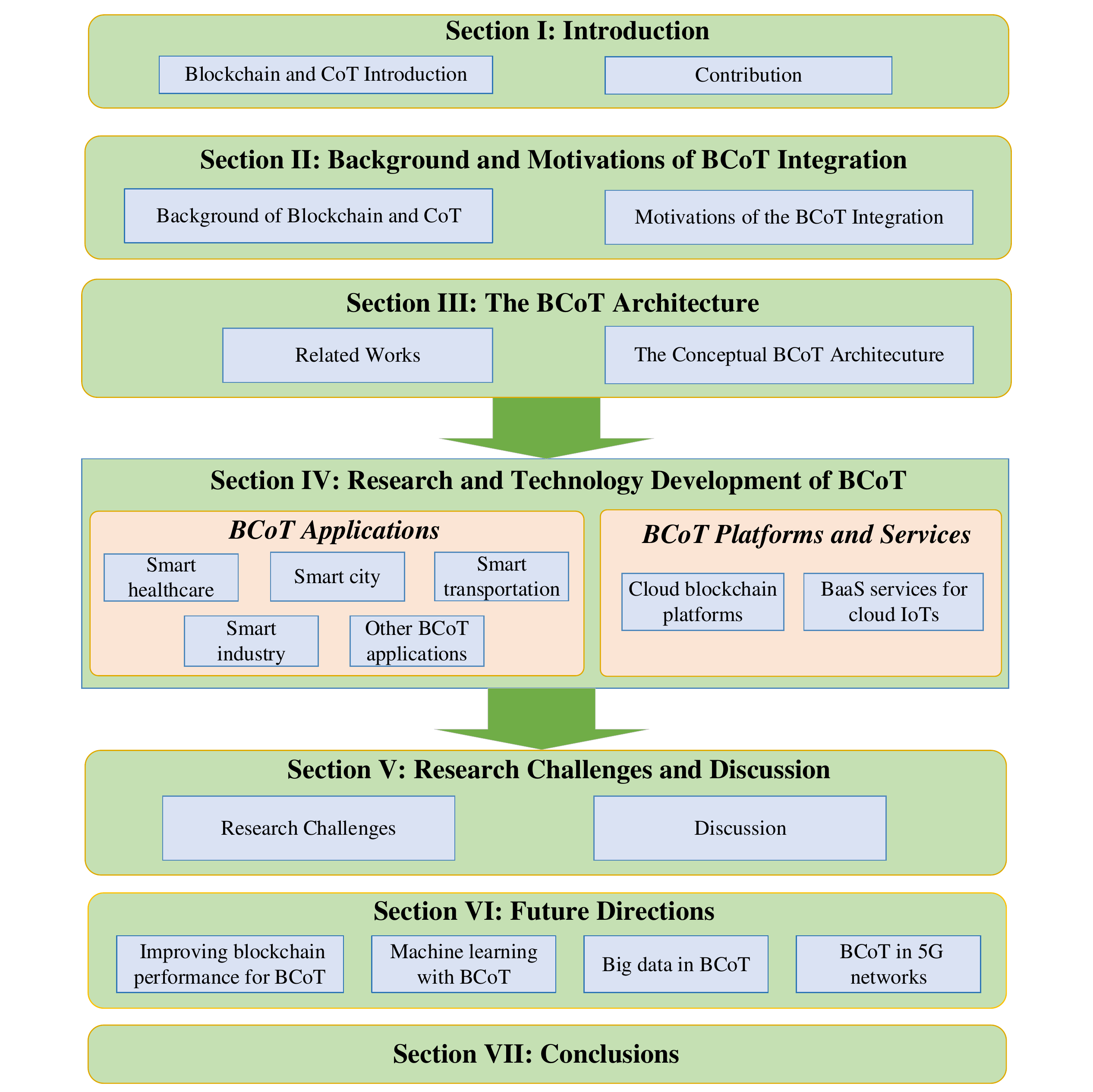}
	\caption{{Organization of the paper.} }
	\label{structure}
	\vspace{-0.17in}
\end{figure*}
 Reviewing the state of the art in the field, we find that BCoT attracts enormous interests of research communities as shown in Fig.~\ref{researchtrend}. Cloud computing and IoT have gained popularity in the last five years with considerable research publications. Interestingly, blockchain is increasingly becoming a hot research area in recent years with a fast-growing research trend, showing a really promising topic for both academics and industries in the future. The sustainable development of CoT and blockchain will drive breakthrough innovations to empower intelligent services and applications. 	
\subsection{	Related Works and Contributions of This Survey}
	
	\textcolor{black}{Many studies in CoT, blockchain and related issues have been investigated over the recent years in a wide range of technical aspects. Many efforts have been made to provide review articles on this research area in different scopes. The survey papers \cite{23}, \cite{24}, \cite{25} presented the review of recent efforts in the adoption of blockchain technology in various IoT scenarios and applications. The authors in \cite{257} also discussed the integration of blockchain technology with IoT. The key focus of this work is on the investigation of the potential of blockchain for IoT applications, from smart manufacturing to the Internet of vehicles, unmanned aerial vehicles, and 5G networks. The survey in \cite{258} paid attention to analysis on the technical aspects of blockchain, such as underlying concepts, networking and consensus strategies. Meanwhile, the authors in \cite{26} discussed research issues, challenges and opportunities of combination between blockchain and cloud computing. The work \cite{27} presented a survey on the use of the blockchain technology to provide security services and its technical properties to solve associated challenges in various application domains, including IoT and Cloud computing. More recently, the overview of the integrated model of blockchain and edge computing, an extended cloud computing concept, was discussed in the survey \cite{28}. Table I summarizes the main topics and contributions of the literature surveys related to BCoT and our paper. }
	
	\textcolor{black}{\begin{table}[ht]
		\centering
		\caption{\textcolor{black}{Surveys on blockchain and Cloud, IoT technologies.} }
		\label{table1}		
		\setlength{\tabcolsep}{5pt}
		\begin{tabular}{|c|P{2cm}|p{4.4cm}|}
			\hline
			\textbf{Paper}& \centering
			\textbf{Topic}&	
			\textbf{Main contributions}
			\\
			\hline
			[23]&	Blockchain and IoT	&A survey of blockchain protocols for IoT, research issues and challenges of IoT-blockchain integration.
			\\
			\hline
			[24]&	Blockchain and IoT&	A comprehensive survey of underlying blockchain concepts, architectures, applications for IoT.	
			\\
			\hline
			[25]&	Blockchain and IoT&	A brief review on the usage of blockchain for IoT.
			\\
			\hline
			[26]&	\textcolor{black}{Blockchain and IoT}&	\textcolor{black}{The investigation of the potential of blockchain for IoT applications.}
			\\
			\hline
			[27]&	\textcolor{black}{Blockchain and IoT}&	\textcolor{black}{Analysis on the technical aspects of blockchain and potentials of blockchain for IoT.} 
			\\
			\hline
			[28]&	Blockchain and Cloud&	A brief survey of blockchain in cloud computing and its security solutions for cloud-based applications. 
			\\
			\hline
			[29]&	Blockchain and Cloud&	An introduction of blockchain for cloud platforms with associated challenges opportunities. 
			\\
			\hline
			[30]&	Blockchain and edge computing&	A systematic survey of the combination of blockchain and edge computing
			\\
			\hline
			\textit{This paper}& 	Blockchain and CoT&	A comprehensive review on the integration of blockchain and CoT with a detailed discussion on concepts, architectures, BCoT applications and BCoT platforms along with challenges and future research directions. 
			\\
			\hline
		\end{tabular}
	\end{table}}

	\textcolor{black}{Although blockchain and CoT have been studied extensively in the literature works, there is no existing work to provide a comprehensive survey on the combination of these important research areas, to our best knowledge. In comparison to the above review works, in this paper, we provide an extensive survey on the integration of blockchain and CoT with a comprehensive discussion on many aspects, ranging from concept background, integrated architectures to application domains, service platforms and research challenges.} The main goal of this survey is to provide readers with thorough knowledge of the blockchain and CoT integration which is collected from respective websites, technical reports, academic articles and newspapers. 
	{The main contributions of this survey are highlighted as follows.}
	
	{\begin{enumerate}
		\item We provide a state-of-the-art survey on the corporation of blockchain and CoT with a comprehensive discussion on different technical aspects, from BCoT background, integration motivations to the conceptual BCoT integrated architecture. 
		\item We present the updated review on the use of BCoT models in various application domains. We analyse the benefits of BCoT adoption and the important lessons learnt in each use case are then summarized. Moreover, the emerging BCoT platforms and services are also presented and discussed. 
		\item From the extensive review on BCoT integrations, we identify possible research challenges and open issues in the field. Some future research directions are also explored to extend the scope of BCoT in future services and applications.
	\end{enumerate}	}

\subsection{Structure of The Survey}
The structure of this survey is organized as Fig.~\ref{structure}. Section II describes the background knowledge of both blockchain and CoT and presents the motivations of the BCoT integration. The conceptual BCoT architecture is presented in Section III. We review the recent developments of BCoT in Section IV with extensive discussions on the roles of BCoT in a wide range of application domains. Moreover, the emerging BCoT platforms and services are also presented and discussed. Section V discusses possible research challenges and open issues in the BCoT integration, while future directions are outlined in Section VI. Finally, Section VII concludes the paper.

\section{{Blockchain, CoT, and Integration Motivation}}
In this section, we first introduce the background knowledge of blockchain and CoT. Then, we present the motivations of the integration of such two technologies. 

\subsection{{Blockchain and Cloud of Things}}
\subsubsection{{Blockchain}}

\begin{figure}
	\centering
	\includegraphics [height=4.3cm,width=8.8cm]{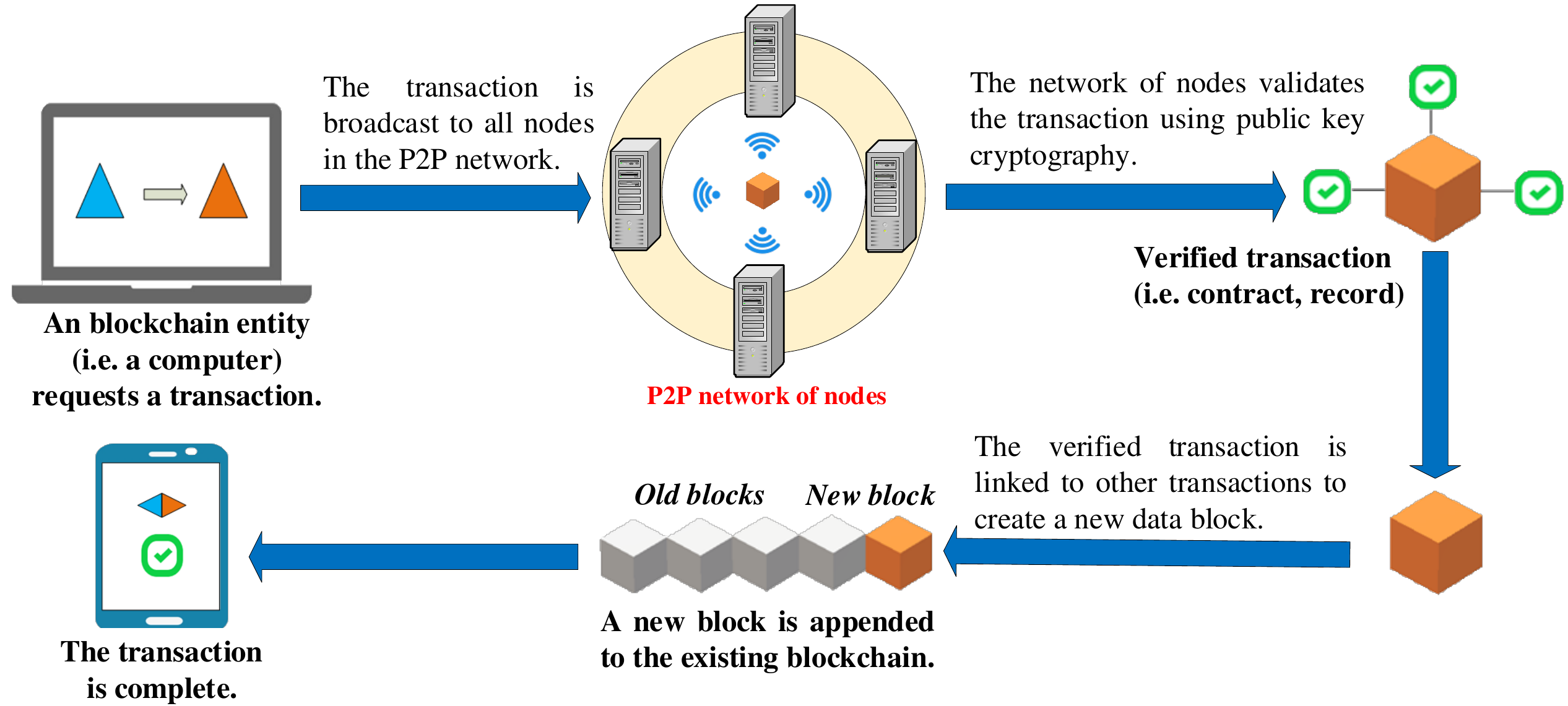}
	\caption{The concept of blockchain operation.}
	\label{blockchainconcept}
	\vspace{-0.15in}
\end{figure}

Blockchain is mostly known as the technology underlying the virtual cryptocurrency Bitcoin which was invented in 2008 by a person known as Satoshi Nakamoto [8]. In a nutshell, the blockchain is briefly explained as public, trusted and shared ledger based on a peer-to-peer network. This emerging technology has also recently become a hot topic for researchers and been argued to innovate blockchain-based applications beyond Bitcoin. The core idea of the blockchain network is decentralization which means blockchain is distributed over a network of nodes. Each node has the possibility of verifying the actions of other entities in the network, as well as the capability to create, authenticate and validate the new transaction to be recorded in the blockchain. This decentralized architecture ensures robust and secure operations on blockchain with the advantages of tamper resistance and no single-point failure vulnerabilities. Basically, blockchain can be classified into two main categories, including public (or permission-less) and private (or permissioned) blockchain. A public blockchain (e.g., Bitcoin platform) is an open network which means it is accessible for everyone to join and make transactions as well as participate in the consensus process. Meanwhile, private blockchain is an invitation-only network managed by a central entity and all participations in blockchain for submitting or writing transactions have to be permissioned by a validation mechanism. A general concept on how blockchain operates is presented in Fig.~\ref{blockchainconcept} and the most popular and promising blockchain platforms are summarized in Table II. 
	\begin{table*}[ht]
	\centering
	\caption{Popular blockchain platforms. }
	\label{table}
	
	\setlength{\tabcolsep}{5pt}
	\begin{tabular}{|c|c|c|P{2cm}|P{2cm}|c|c|}
		\hline
		\textbf{Platforms}& 
		\textbf{Consensus}&	
		\textbf{Operation Modes}& 
		\centering \textbf{Smart contract support?}&
		\centering \textbf{Programming language}& 
		\textbf{Latest version}&
		\textbf{Open source?} 
		\\
		\hline
		Bitcoin&PoW&Public &\centering Yes&\centering Ivy, RSK, BitML&v0.18.0, May. 2019&Yes \cite{47}
		\\
		\hline
		Ethereum&	PoW, PoS&	Public and permissioned	&\centering Yes&	Solidity, Flint, SCILLA &v1.8.27, Apr. 2019&	Yes \cite{48}
		\\
		\hline
		Hyperledger &	PBFT	& Permissioned &\centering 	Yes &\centering 	Go, Node.js, Java&	v2.0 Alpha, Apr. 2019&	Yes \cite{49}
		\\
		\hline
		IBM Blockchain&	PoW, PoS&	Permissioned	&\centering Yes &\centering 	Go, Java&	v2.0, Jun. 2019	&Yes \cite{50}
		\\
		\hline
		Multichain&	PBFT&	Permissioned	&\centering No&	\centering C++, Go, Java, Python, PHP& 	beta 2.0, Mar. 2016&	Yes \cite{51}
		\\
		\hline
		Hydrachain&	PoW, PoS&	Permissioned&\centering 	Yes&\centering 	Python&	hydrachain 0.3.2, 2018&	Yes \cite{52}
		\\
		\hline
		Ripple&	PoW&	Permissioned&\centering 	Yes&\centering 	C++&	v1.2.4, Apr. 2018	&Yes \cite{53}
		\\	
		\hline
		R3 Corda&	PoW, PoS	&Permissioned&\centering 	Yes &\centering 	Kotlin, Java&	V4.0, Feb. 2019&	Yes \cite{54}
		\\	
		\hline
		BigChainDB	&BFT&	Public and permissioned	&\centering Yes&\centering 	Java&	v2.0.0b9, Nov. 2018&	Yes \cite{55}
		\\
		\hline
		Openchain&	Partionned &{Permissioned}&	\centering Yes	&\centering 	C++, Java&	v0.7.0, Nov. 2017&	Yes \cite{56}
		\\
		\hline
	\end{tabular}
\end{table*}

{A blockchain network is built from some key components, including data block, distributed ledger, consensus, and smart contracts. To be clear, each block contains a number of transactions and is linked to its immediately-previous block through a hash label. In this way, all blocks in the chain can be traced back to the previous one, and no modification or alternation to block data is possible \cite{39}. A distributed ledger is a type of database which is shared and replicated among the entities of a peer-to-peer network. Moreover, blockchain consensus is a process used to reach agreement on a single data block among multiple unreliable nodes, aiming to ensure security in a blockchain network. Final, smart contracts are programmable applications that run on a blockchain network according to predefined contractual conditions such as payment terms, liens, confidentiality, and even enforcement \cite{44}. }

Blockchain can provide high-security properties for its applied scenarios, such as CoT. The most important feature is decentralization which means blockchain does not rely on a central point of control to manage transactions. This exceptional property brings promising benefits, including eliminating single point failure risks due to the disruption of central authority, saving operational costs and enhancing trustworthiness. Further, blockchain is able to keep transaction data immutable over time. The hashing process of a new block always contains metadata of the hash value of the previous block, which makes the chain strongly unalterable. In this way, it is impossible to modify, change or delete data of the block after it is validated and placed in the blockchain. Another important feature is transparency which stems from the fact that all information of transactions on blockchain is viewable to all network participants. In other words, the same copy of records of blockchain spreads across a large network for public verifiability. As a result, all blockchain users can fully access, verify and track transaction activities over the network with equal rights.

{In addition to these security benefits, blockchain remains some problems that need to be considered to make it well suitable for integrating with CoT. For example, how to achieve energy efficiency for blockchain in BCoT applications is an important issue. The decentralized consensus algorithms in blockchains often require extensive processing power and high computing energy to mine blocks and maintain the blockchain network. This makes blockchain infeasible to resource-constrained IoT devices in CoT applications. Although we can perform energy-intensive blockchain mining in a centralized cloud, this would essentially negate the advantages of a distributed CoT system. Another issue is the throughput of blockchain systems. In fact, blockchain has much lower throughput in comparison to non-blockchain applications. For example, Bitcoin is able to execute a maximum of only four transactions/second, and the  throughput of Ethereum achieved is about 20 transactions/second, while Visa can process up to 1667 transactions/second [24]. Clearly, the existing blockchain networks still remain scalability bottlenecks in terms of the number of replicas and limited throughput. Many current blockchain systems suffer from high block generation time which in turn reduces the overall blockchain throughput.  Further, if all transactions are stored in a chain, the blockchain size will become very large \cite{70}. Considering real-world CoT scenarios, e.g., smart cities, the IoT data volume is enormous and thus will result in a rapid growth in the IoT blockchain size. These factors thus should be taken into account in designing blockchain platforms for sustainable BCoT applications. }
\subsubsection{Cloud of Things}
\begin{figure}
	\centering
	\includegraphics [height=4cm,width=8cm]{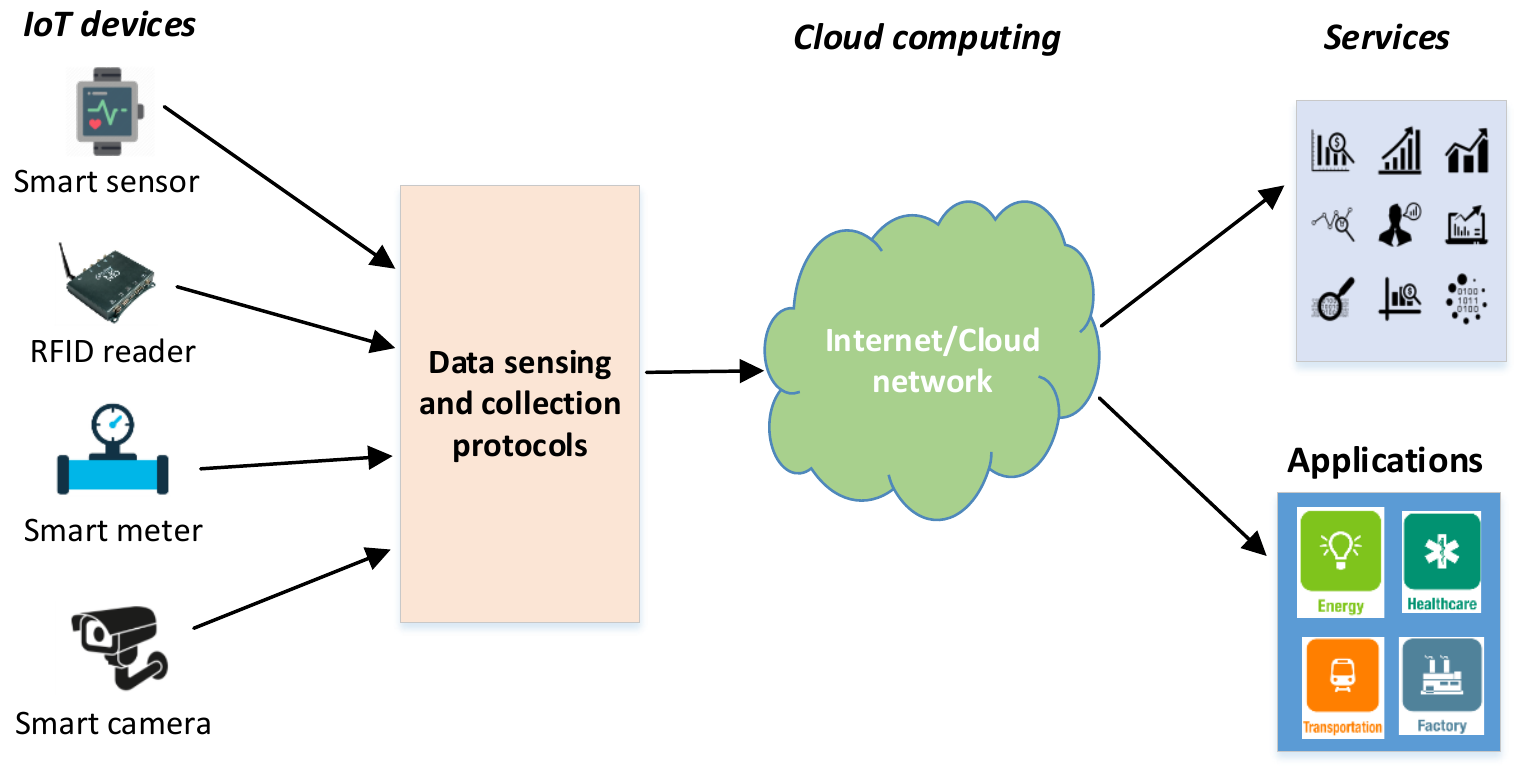}
	\caption{The general concept of CoT.}
	\label{CoTconcept}
	\vspace{-0.1in}
\end{figure}

Nowadays, IoT has constituted a fundamental part of the future Internet and drawn increasing attention from academics and industries thanks to its great potentials to deliver exciting services across various applications. IoT seamlessly interconnects heterogeneous devices and objects to create a physical environment where sensing, processing and communication processes are implemented automatically without human involvement. However, massive volumes of data generated from a large number of devices in current IoT systems become a bottleneck in guaranteeing the desired Quality of Service (QoS) because of constrained power and storage resources of IoT devices. Meanwhile, cloud computing has unlimited resources in terms of storage and computation power, which can provide on-demand, powerful and efficient services for IoT use domains. Especially, the convergence of cloud computing with IoT paves the way for a new paradigm as CoT, which can empower both worlds. Indeed, the wealth of resources available on the cloud is highly beneficial to IoT systems, while cloud can gain more popularity in real-life applications from integrating with IoT platforms [61]. Moreover, CoT can transform current IoT service provision models with minimal management effort, high system performance and service availability. The general concept of CoT is shown in Fig.~\ref{CoTconcept} with a network architecture of IoT devices, cloud computing, analytic services and application layer. In this hierarchy, IoT devices are used to sense and collect data from local environments. However, due to their limited computing resources, IoT devices will transmit recorded data to the cloud for data acquisition. Cloud computing can provide a powerful capability of data processing and storage. Analytic services can be provided to support IoT systems, such as historic data monitoring, information storage or statistical analysis. The results of cloud data processing are used to serve end applications, aiming to facilitate IoT service provisions and meet requirements of end users. 

{In general, the CoT platform can offer instant services to users anywhere and anytime thanks to automatic resource provision capabilities of cloud computing. It enables autonomous service delivery without the need for human engagement. With unlimited virtual processing capabilities of cloud computing, CoT open up new opportunities to enhance IoT computation by enabling data offloading and executing data remotely. This not only improves computation abilities of local devices, but also addresses effectively issues of IoT systems in terms of energy-saving and bandwidth preservation. Importantly, CoT can offer simplified and automatic IT maintenance and management solutions by exploiting cloud servers, virtual machines and resource infrastructure. IoT users can interact easily with cloud computing to implement functionalities without any requirement for software installation as well as human involvement. Moreover, the provision of system management models available on clouds also supports well boundless communications and interconnections between IoT devices together, between things and users to empower ubiquitous applications, which would promote the comprehensive collaborations of multiple IoT ecosystems in the future Internet. }

\subsection{Motivation of the Integration of Blockchain and CoT}

In this subsection, we highlight the motivation of the integration which comes from the security challenges of CoT, technical limitations of blockchain and the promising opportunities brought by the incorporation of such two technologies. 

\subsubsection{Security Challenges in CoT}

CoT support ubiquitous computing services with large data storage and high system performance, but still remains some critical challenges \cite{62}, \cite{63} as below. 

\textit{Data availability:} In the current cloud network architectures, cloud services are provided and managed centrally by the centralized authority. However, this configuration is vulnerable to single-point failures, which bring threats to the availability of cloud services for on-demand IoT access. A centralized cloud IoT system does not guarantee seamless provisions of IoT services when multiple users request simultaneously data or cloud servers are disrupted due to software bugs or cyber-attacks.

\textit{Privacy management:} Although the centralized cloud IoT can provide convenient services, this paradigm raises critical data privacy concerns, considering a large amount of IoT data being collected, transferred, stored and used on the dynamic cloud networks. In fact, IoT users often place their trust in cloud providers managing the applications while knowing very little about how data is transmitted and who is currently using their information \cite{64}. Even in the distributed cloud IoT paradigms with multiple clouds, IoT data are not fully distributed but stored in some cloud data centres at high density \cite{65}. In this context, IoT data may be leaked if one of the cloud servers is attacked.   

\textit{Data integrity: }The storage and analysis of IoT data on clouds may give rise to integrity concerns.  Indeed, due to having to place trust on the centralized cloud providers, outsourced data is put at risks of being modified or deleted by third parties without user consent. Moreover, adversaries can tamper with cloud data resources for financial or political purposes \cite{67}, all of which can breach data integrity.  For these reasons, many solutions using public verification schemes based on a third-party auditor have been proposed, but they potentially raise several issues, including irresponsible verification to generate bias data integrity results or invalidated verification due to malicious auditors. Therefore, developing new solutions to solve efficiently data integrity challenges is vitally necessary for CoT systems.

\subsubsection{Technical Limitations of Blockchain}

Although blockchain has its unique promise to disrupt services like CoT, it still remains several critical challenges in its development in terms of complexity, and security flaw.

\textit{Complexity:} In IoT networks, in order to implement validation on transactions, IoT devices act as blockchain participants to run the consensus process to solve complex mathematical puzzles, which requires powerful computation hardware. Unfortunately, this is challenging to meet such requirements due to the constraints of IoT resources. Even in the case of IoT devices with relatively high computing capacities, running complex blockchain process may require intensive resources involving electricity and human management. This would raise concerns of users about high operational costs which would hinder wide deployment of blockchain-based systems.

\textit{Security flaw:} The final limitation of current blockchains may be an unavoidable security flaw. If more than half of computers working as blockchain nodes to control computing power, attackers may modify consensus architectures and prevent new transactions from obtaining confirmations for malicious access. This is also called as 51\% attack which is highlighted in the Bitcoin concept. Without having a comprehensive transaction management, blockchain can be put at risks of data breach and system damage. 

In short, the decentralization of CoT lays the ground for blockchain as data security and privacy solutions, and blockchain can leverage the cloud resources in CoT for intensive mining computations and reliable data storage. 

\subsubsection{The Opportunities of Integration of Blockchain and CoT}

Based on complementary roles of blockchain and CoT as well as their potential advantages, the incorporation of such disruptive technologies opens up a wide range of BCoT opportunities, as summarized in the following. 

\textit{Decentralization management:} Motivated by the fully decentralized nature of blockchain, it is possible to build a decentralized BCoT management architecture under the distributed control of peer-to-peer network of cloud nodes and IoT devices. All blockchain peers maintain identical replicas of the ledger data records through decentralized consensus, and trustfulness is shared and distributed equally among the network entities. This decentralized structure eliminates totally single point failure bottlenecks, prevents efficiently disruption of BCoT services, and enhances significantly data availability. 

\textit{Improved data privacy:} The dynamic process of outsourcing IoT data to clouds and data exchange between cloud providers and IoT users are vulnerable to information disclosure and attacks caused by adversaries or third parties. Blockchain with its immutability, integrity and transparency properties is highly suitable for data protection in CoT networks. In fact, to launch a data modification attack in a BCoT system, an adversary would try to modify the records or alter data placed in blockchain. However, this is nearly impossible in practical scenarios where blockchain is preserved and controlled by secure and immutable consensus mechanisms. As a result, properties inherent to blockchain can significantly enhance data privacy for BCoT applications. 

\textit{Improved system security:} Blockchain can provide solutions to improve security for CoT, through the ability to offer important security properties such as confidentiality and availability inherent in blockchain. Indeed, in BCoT networks, all records on the blockchain are cryptographically hashed and transactions are signed by participants so that all user interactions with clouds remain confidential under blockchain-enabled signatures. Furthermore, with the decentralization feature inherent in blockchain, data is replicated across all network members with no single of failure bottlenecks, and thus BCoT promises to provide improved availability. Specially, the resourceful cloud computing can provide off-chain storage solutions to support data availability of the on-chain storage mechanisms once the main BCoT network is interrupted due to external attacks. On the other side, the implementation of blockchain algorithms on clouds may enhance security of the blockchain system itself. For example, clouds can use their available network security tools to maintain and preserve blockchain software, e.g., mining mechanism, against potential threats. The advantage of cloud computing for blockchain is proven through recent successful integration cloud-blockchain projects, such as Oracle blockchain (2017) and iExec blockchain (2018) projects \cite{72}.  

\textit{Reduced system complexity:} By integrating blockchain with cloud computing, BCoT can reduce significantly complexity of system implementations. This integration is known as blockchain-as-a-service, where well-defined platforms are available to set up and run blockchain for BCoT projects without worrying about underlying hardware technologies \cite{73}. Moreover, blockchain algorithms now can be run online using cloud infrastructure, which is promising to reduce resource costs for running blockchain. Obviously, the convergence of blockchain and CoT opens up various opportunities to accelerate BCoT deployments on the large scale with simple and cheap implementations.

\subsubsection{Feasibility of the BCoT Integration}

At present, more and more large companies implement BaaS projects to assess the feasibility of the integration of blockchain in cloud computing. Large companies such as Amazon, Microsoft, IBM, Oracle have responded by launching BaaS platforms for IoT on cloud computing. The Amazon cloud provider develops a BaaS platform \cite{188} for IoT business models. For example, an IoT healthcare system was developed in \cite{105} by using Amazon BaaS architecture. In this project, the Ethereum blockchain platform hosted on Amazon cloud helps to implement a health data sharing framework on mobile clouds with high security and privacy. Moreover, IBM cloud \cite{187} also introduces a well-developed BaaS platform for IoT users. The platform has been showcased in a vehicular network [206]. In this project, the IBM IoT platform is integrated with IBM BaaS services to manage vehicle sensor data (vehicle-to-vehicle messages and vehicle monitoring data) and ensure security during data sharing within the vehicular network. Meanwhile, the BaaS platform of Oracle cloud \cite{189} has proven its great potentials through a wide range of BCoT projects, such as banking, healthcare data management, and payment industry. Such above examples have shown high feasibility of the blockchain adoption in cloud computing for solving complex issues in terms of security, network performance for IoT applications. We will detail the development of BaaS models in various IoT domains in the following sections.

\section{	The Architecture of Integration of Blockchain and CoT}

In this section, we review thoroughly the literature studies towards the integrated BCoT models of blockchain and CoT. We then propose a conceptual BCoT architecture with the fundamental concept and basic ideas of the integration which would be applicable to various scenarios.

\subsection{Related Works}

With the current growing interest in the blockchain and CoT, many new integrated BCoT platforms and systems have been proposed in the literature studies to provide security solutions and applications \cite{74}, \cite{75}, \cite{76}, \cite{77}, \cite{78}. The study \cite{79} proposed a cloud-centric IoT framework enabled by smart contracts and blockchain for secure data provenance. Blockchain incorporates in cloud computing to build a comprehensive security network where IoT metadata (e.g., cryptographic hash) is stored in blockchain while actual data is kept in cloud storage, which makes it highly scalable for dense IoT deployments. Another work in \cite{80} introduced a blockchain-cloud network for access control with four main components: IoT devices, a data owner, a blockchain network and a cloud computing platform. Similarly, a hierarchical access control structure for BCoT was investigated in \cite{81}. The blockchain network topology involves distributed side blockchains deployed at fog nodes and a multi-blockchain operated in the cloud, which would speed up access verification offer flexible storage for scalable IoT networks. In addition, to protect BCoT in security-critical applications, a forensic investigation framework is proposed using decentralized blockchain \cite{82}. 
Following by the advantages of BCoT conjunction, \cite{83}, \cite{84} provided secure identity management solutions which allow cloud service providers to autonomously control and authenticate user identity in BCoT. Blockchain is combined with virtual clouds to support identity verification in a fashion there is no prior requirements on trust between cloud users and cloud providers.  On the other side, data management is also critical in interconnected CoT where IoT data is enormous and thus requires careful management for data privacy objectives. Motivated by this, the work \cite{85} presented a blockchain-based data protection mechanism which can prevent effectively inappropriate IoT data movement due to malicious tampering during Virtual Machine (VM) migration on cloud computing. Also, a Mchain construction method is applied to integrity evaluation on VM measurements data \cite{86}. In this architecture, a two-layer blockchain network, which includes a data validation layer and a PoW task layer, is integrated with IaaS cloud to enhance system integrity \cite{87}.

{ Moreover, the work in \cite{300} also considered an integrated blockchain-CoT architecture where the focus was on solving the mining issue by offloading mining tasks to cloud nodes from IoT devices. Then, a joint problem of user access association and cloud resource allocation is formulated that is then solved by deep reinforcement learning (DRL). In the same direction, the authors in \cite{301}, \cite{302} also considered the offloading issue in BCoT networks, in order to optimize the economic cost of IoT devices. The study in \cite{303} paid attention to the cloud service quality in the BCoT systems. In this case, the blockchain plays an important role in providing trust and reliability for high-quality cloud service provisions. The combination of cloud computing with blockchain was also considered in \cite{304}. Here, the computing resources of remote cloud are allocated at the network edge to provide low-latency and real-time computing services for IoT devices. Meanwhile, the resource management in BCoT systems was studied in \cite{305} where the blockchain is able to preserve data privacy during the resource trading between cloud providers and IoT users. }

In general, most of the above BCoT platforms are based on a single cloud and may be enough for some applications. However, with complex IoT systems which require huge network resources to serve numerous IoT users, inter-cloud BCoT integration would be more efficient and convenient \cite{20}. As a result, BCoT architectures have been extended to multi-cloud models for complex collaborative scenarios \cite{88}, \cite{89}. As an example, a BCoT framework was proposed in a joint cloud collaboration environment where multiple clouds are interconnected securely by a peer-to-peer ledge network \cite{90}.  Further, the single cloud can offer instant services for IoT users via blockchain which also mitigates risks of malicious attacks \cite{91}. Moreover, \cite{65} proposed a cloud federation model which enable distributed resource provisions using an individual cloud under the management of blockchain network. Besides, a BCoT model with micro-clouds was introduced by \cite{92} using blockchain-enabled distributed ledgers.
\begin{figure*}
	\centering
	\includegraphics[height=16cm, width=18cm]{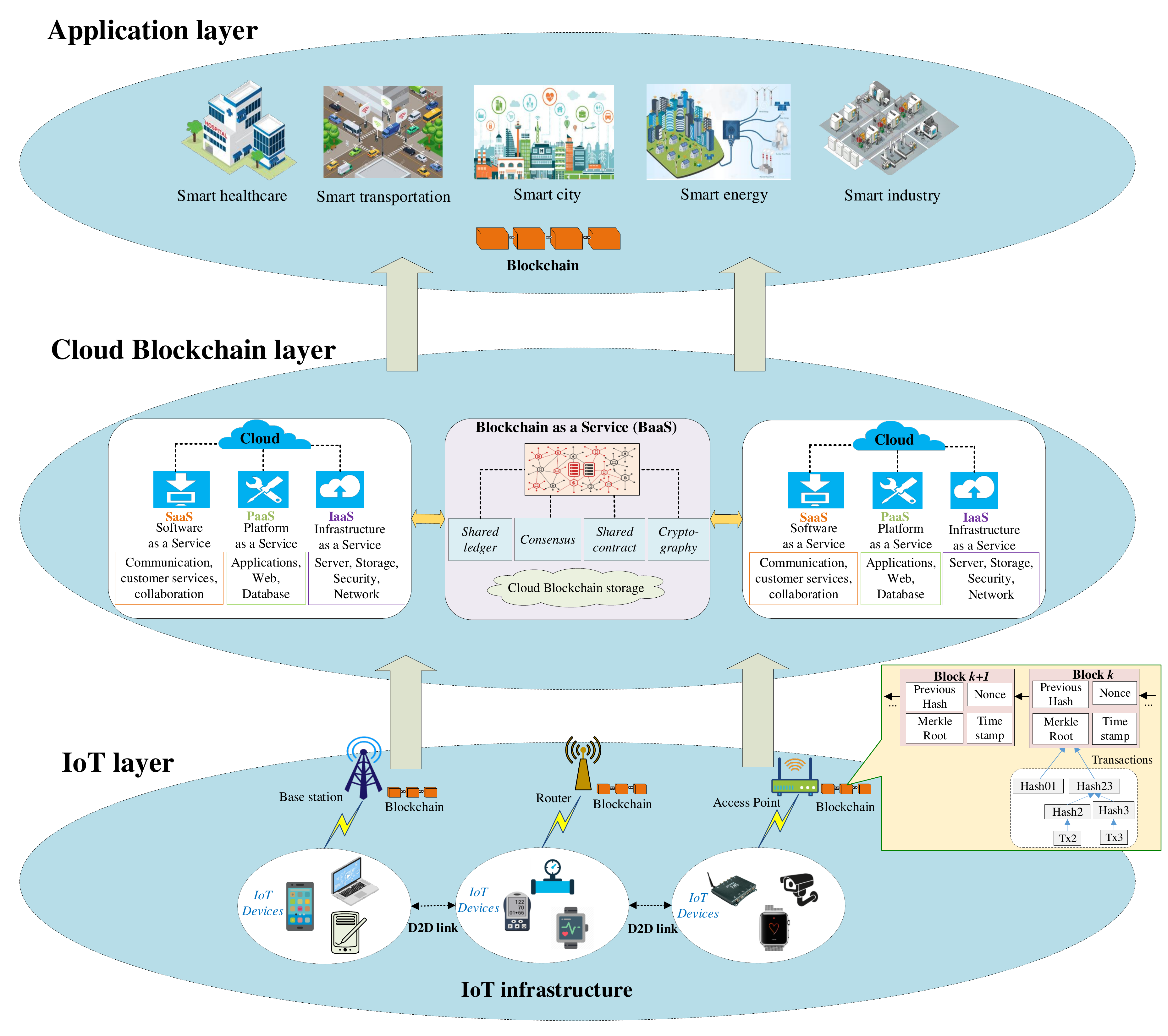}
	\caption{The conceptual BCoT architecture.   }
	\label{BCoTarchitecture}
	\vspace{-0.17in}
\end{figure*}

\subsection{The Conceptual BCoT Architecture}

Motivated by extensive literature review, we propose a conceptual BCoT architecture as shown in Fig.~\ref{BCoTarchitecture}, including three main layers: IoT layer, cloud blockchain layer and application layer. Details of each layer and the general concept will be presented as the following.

\subsubsection{IoT Layer}

IoT devices are responsible for harvesting data from local environments and transmitting wirelessly it to nearby gateways such as base station, router or wireless access point. An IoT device holds a blockchain account (like a wallet in Bitcoin) which allows it to join the blockchain network to perform transactions (e.g., offloading data) and interactions with cloud services. \textcolor{black}{Specially, each resource-limited IoT device (e.g., a wearable sensor) may act as a lightweight node that can participate in the validation process of a transaction through its representative gateway. It is feasible in blockchain-based sensor network scenarios such as [105], [209], [214] where small sensors are connected with blockchain via its gateway (e.g., a smartphone or a fog node). All interactions of sensors with blockchain such as creating transactions, offloading data or even mining tasks, are performed by the gateway \cite{79}.} Meanwhile, for IoT devices with relatively large resources such as computers or powerful smartphones, they have enough capacities to serve other lightweight IoT sensors and maintain the full blockchain. IoT devices can also interact each other through IoT gateways to achieve corporative communication (e.g., device to device (D2D) communication in collaborative networks). Such a hybrid communication concept offers highly flexible services for IoT users in a secure and efficient manner.

\subsubsection{Cloud Blockchain Layer}

This plays as a middleware between the IoT network and industrial applications in the BCoT architecture. For a generic architecture, we pay attention to a blockchain platform with multiple clouds, but it also reflects comprehensively technical aspects of a single-cloud BCoT architecture.  This model exhibits two merits: 1) ensuring highly secure network management via blockchain and 2) providing on-demand and reliable computing services for large-scale IoT applications. The integrated cloud blockchain layer consists of blockchain services and cloud computing services.

- \textit{Blockchain services:} The main purpose of blockchain in the proposed architecture is to provide secure network management. The blockchain network is deployed and hosted on a cloud platform as Blockchain as a Service (BaaS).  In particular, BaaS can offer a number of blockchain-enabled services to support IoT applications.

\begin{itemize}
	\item Shared ledger: It represents the database that is shared and distributed among BCoT members (e.g., IoT users, cloud nodes and blockchain entities). The shared ledger records transactions, such as information exchange or data sharing among IoT devices and cloud. It enables industrial networks where cloud users can control and verify their own transactions when communicating with blockchain cloud.
	\item Consensus: It provides verification services on user transactions by using consensus mechanisms such as PoW, PoS run by a network of miners. This service is highly necessary for BCoT in improving blockchain consistency and ensuring high security for the system. Interestingly, IoT users can use their virtual cloud machines to join the consensus process in order to receive rewards as a result of their efforts (e.g., cryptocurrency in Bitcoin).
	\item 	Shared contract: BCoT also offer smart contract services to applications. With its self-executing and independent features, smart contracts are highly beneficial to build business logic and trust in the BCoT system. Furthermore, smart contracts provide security services on user access authentication or data sharing verification once the IoT peer nodes perform transactions, which also supports to maintain security over the cloud blockchain. 
	\item	Cryptography: This is responsible for providing public-key cryptography to secure all information and storage of data among IoT and cloud entities. Digital signatures ensure any data being recorded in blockchain is true and untampered with, and this improves immutability and security for user transactions.
\end{itemize}
In addition to such services, BaaS also offers cloud blockchain storage. The decentralized cloud storage based on blockchain can be built on the cloud platform. Blockchain-based storage manages IoT data through its hash values and implements verification periodically to detect any data modification potentials. For example, InterPlanetary File System (IPFS) \cite{93} is a blockchain-based storage system which is now available on cloud, allowing to store securely among storage nodes. This has also been proven to solve effectively data storage issues brought by centralized cloud models in terms of data leakage and storage management.

- \textit{Cloud computing services: }In the BCoT architecture, cloud computing uses its full services to support applications, including Software as a Service (SaaS), Infrastructure as a Service (IaaS) and Platform as a Service (PaaS). Data aggregated by IoT gateways will be received by cloud servers and kept in the cloud blockchain storage. The cloud server also offers intelligent services on offloaded IoT data using available tools such as data mining or machine learning. IoT data can be stored off-chain in cloud database or on-chain in blockchain. On the other hand, multiple clouds can be incorporated to implement functionalities such as data sharing or collaborative system management. In this context, as a middle layer, blockchain layer plays an important role in handling and controlling cloud interactions to facilitate cloud service delivery to IoT users and avoid conflicts among clouds. 

\subsubsection{Application Layer}

Many industrial applications can gain benefits from the BCoT integration in different areas where IoT scenarios are involved, like smart healthcare, smart transportation, smart city, smart energy, and smart industry. BCoT not only provides useful services to industrial applications, such as network management and QoS improvement but also guarantees security and privacy properties for applied domains. For example, in smart healthcare, BCoT can support data processing services thanks to computation ability of cloud, which can assist healthcare providers in analysing intelligently patient information for better medical care. In the meantime, network security of healthcare is ensured with blockchain which offers traceability and verification services during the medical data exchange and data processing. The application of BCoT integration and its benefits to IoT use case domains such as smart industry, smart energy, smart transportation will be extensively analyzed in the next section. 
\begin{figure*}
	\centering
	\includegraphics[height=5cm, width=16cm]{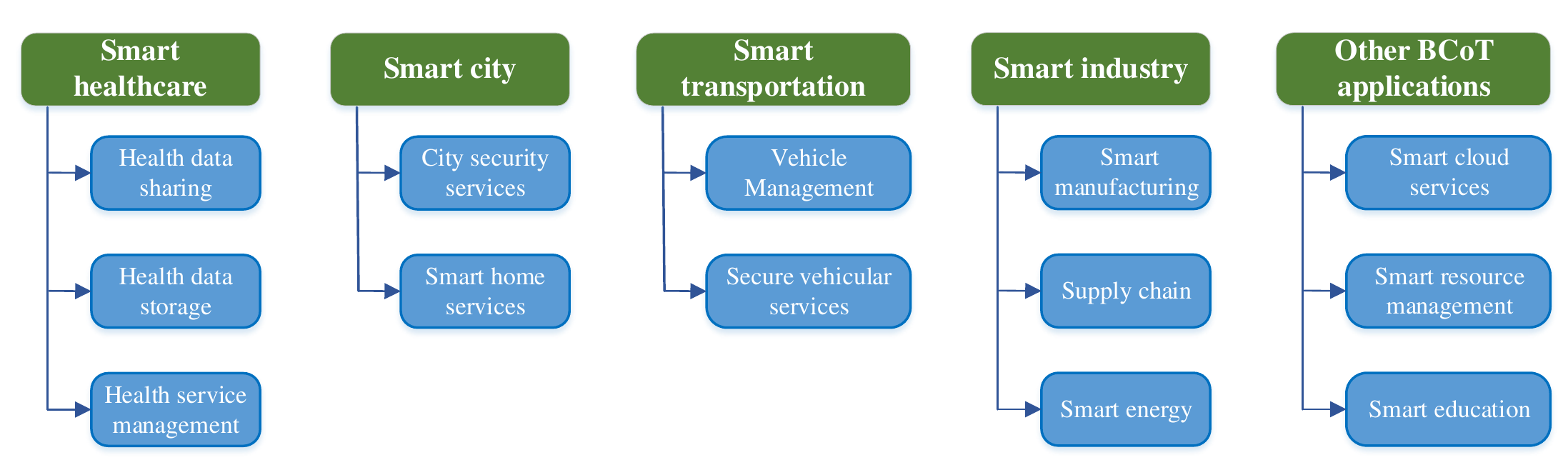}
	\caption{BCoT application domains.  }
	\label{BCoTapplication}
	\vspace{-0.17in}
\end{figure*}

\section{{Research and Technology Development of BCoT}}
In this section, we will present recent advances of BCoT technology. The impacts of BCoT on a wide range of industrial applications are reviewed from the latest research results. Then, we survey recent BCoT developments with platforms, services, and research projects around blockchain and CoT technologies.
\subsection{BCoT Applications}
In this sub-section, we will provide readers with a summary of key BCoT applications across different scenarios, such as smart healthcare, smart city, smart transportation, and smart industry as shown in Fig.~\ref{BCoTapplication}. In particular, we highlight the research findings of the state-of-the-art studies in the use of BCoT paradigms in such applications.  Besides, main lessons learned from the review are also discussed.
\subsubsection{Smart Healthcare}
Healthcare is an industrial sector where organizations and medical institutions provide healthcare services, medical equipment, medical insurance to facilitate healthcare delivery to patients. The adoption of BCoT models can offer great potentials to solve critical issues in terms of security and service efficiency, and thus is possible to advance medical services and transform current healthcare systems \cite{94}, \cite{95}, \cite{96}, \cite{97}. The BCoT integration in healthcare promises to provide new smart services, including efficient health data sharing, healthcare data storage and secure system management, which will be summarized from the literature studies as the following. 

\textit{1.1) Health data sharing} 

CoT enable efficient healthcare data sharing environments where EHRs can be processed and stored online on the cloud storage while users can use their mobile devices (e.g., smartphones) to access their medical information for health monitoring. This promises to offer on-demand healthcare services, save healthcare costs, and improve quality of experience \cite{98}. However, healthcare data sharing based on such dynamic cloud IoT environments is always vulnerable to security and privacy risks due to attack potentials and the lack of trust between healthcare cloud providers, cloud storage, and users. {Blockchain plays a significant role in solving security issues in health data sharing by decentralized data verification of all peers and message validation based on consensus mechanisms. Specially, the traceability of blockchain allows healthcare entities (e.g., healthcare providers, insurance companies, and patients) to trace user access behaviours and detect data attacks, aiming to improve the security of health data sharing in BCoT networks.}

The work \cite{99} introduces a privacy-preserved data sharing scheme which is enabled by the conjunction of a tamper-proof consortium blockchain, cloud storage and medical IoT network. In order to protect medical data, original electrical medical records (EMRs) are stored securely in the cloud under the management of smart contracts while data indexes are kept in blockchain. This ensures that EMRs cannot be modified or altered arbitrarily. A user-centric health data sharing solution is also proposed in \cite{100} using permissioned blockchain on a mobile cloud platform where health data from wearable sensors is synchronized to cloud for data sharing with healthcare providers and insurance institutions. Another work in \cite{101} presents a data sharing framework with fine-grained access control, by combining a decentralized storage system interplanetary file system (IPFS), an Ethereum blockchain, and attribute-based encryption technology. An access control design based on smart contract is also proposed to implement keyword search in decentralized cloud storage for sharing services.

Although BCoT can help achieve secure data sharing, the storage of all health data on blockchain will slow down transaction operations and put sensitive patient information at risks of data leakage and sharing security concerns. Motivated by such challenges, \cite{102} proposes a conceptual scheme for exchanging personal continuous dynamic health data using cloud storage and blockchain. In particular, large health datasets are encrypted and stored as off-the-chain in cloud storage, while only metadata (e.g., hash values) of raw data is kept in blockchain, which would overcome the size limitation of the large data storage in BCoT systems. 

\begin{figure}
	\centering
	\includegraphics[height=4.7cm,width=8cm]{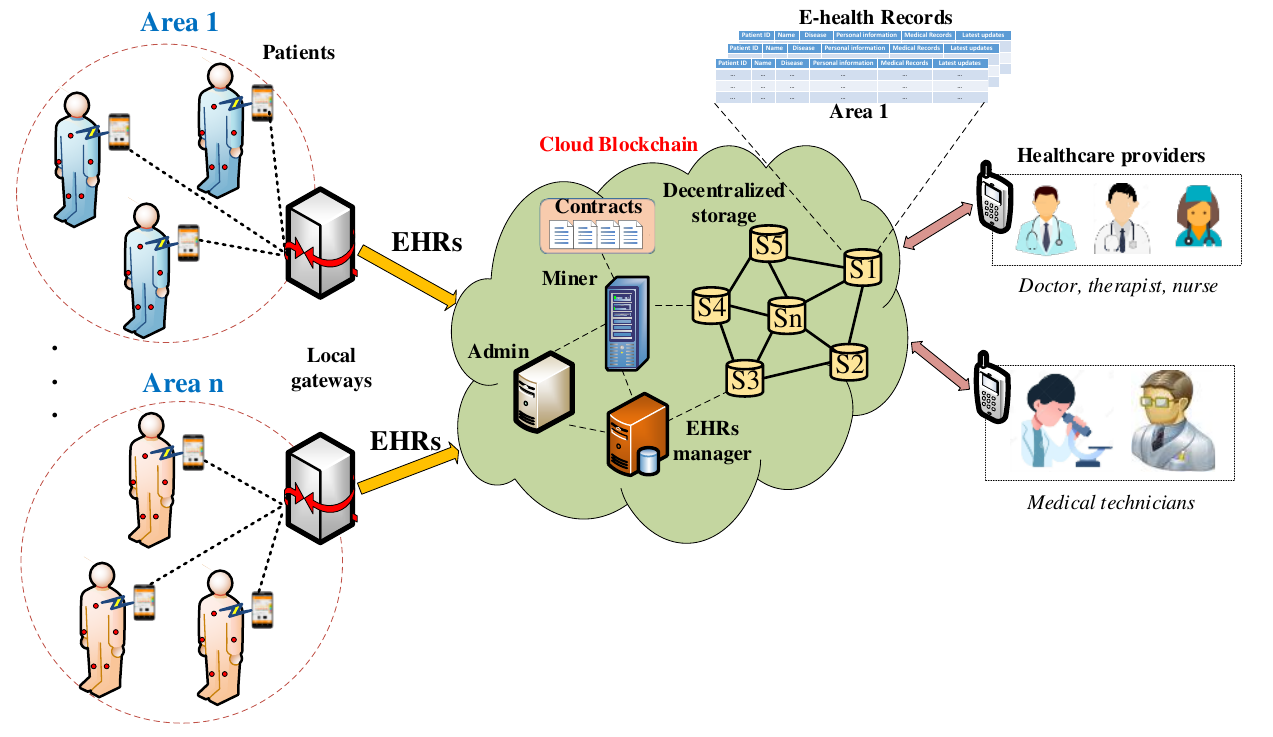}
	\caption{A smart e-health data sharing system \cite{105}.}
	\label{Healthcare}
	\vspace{-0.15in}
\end{figure}

In line of discussion, the authors in \cite{103} propose a trust-less medical data sharing, called MeDShare, to enable data exchange among untrusted cloud service providers (CSP) using the blockchain. The focus is on an access control design based on smart contracts to trace access behaviours of data users, but access control issues associated with sensitive data in the cloud data pool remains unsolved. Therefore, the work \cite{104} proposes secure cryptographic approaches (including encryption and digital signatures) to provide efficient access control which acts as a monitoring system layer to achieve data user authentication for cloud data sharing. More interesting, in our recent work \cite{105}, a mobile cloud blockchain platform is proposed to implement dynamic EHRs sharing where blockchain is integrated with cloud computing to manage user transactions for data access enabled by smart contracts. In particular, a decentralized storage IPFS run by blockchain is combined with cloud computing to make data sharing more efficient in terms of low latency, easy data management and improved data privacy. The concept of the proposed scheme can be seen in Fig.~\ref{Healthcare}.

\textit{1.2) Health data storage}
 
{Blockchain is able to enhance data integrity and traceability for healthcare data storage. Encrypted health records are stored in cloud blockchain under the control of smart contracts. By using blockchain, the system provides the full data integrity to patients and data users. } Vulnerabilities regarding data preservation are addressed effectively by using cryptographic functions along with blockchain, improving integrity, accountability, and security for cloud data storage \cite{106}. For example, the work \cite{107} introduces a secure cloud-based EHR system on blockchain with five entities: key generation centre, hospitals, patients, medical clouds, and data consumers like insurance company. In this network, medical data is stored in the blockchain associated with a complete copy of the timestamp, consequently increasing the integrity and traceability of healthcare records.
The work in \cite{108} integrates the medical data into the infrastructure of cloud blockchain called Blockcloud. Blockchain has distributed ledger where encrypted medicine data transactions are stored on cloud storage as a blockchain entity. Any modifications on medical records in cloud storage will be identified by blockchain via the P2P network.

In \cite{109}, a modified BCoT scheme is proposed for decentralized health data privacy. The architecture consists of overlay network, cloud storage servers, healthcare providers, smart contracts and patients. Specially, blockchain is interconnected with cloud storage via a P2P network where each cloud storage keeps medical records into blocks and the hash value of these blocks is stored in blockchain. This makes any changes in data possible to be easily traced.

\textit{1.3) Healthcare service management}

 The BCoT paradigm may offer unprecedented breakthroughs with new smart medical services, such as decentralized healthcare, secure user management or medical operation control \cite{110}. {Blockchain can be used to build a secure healthcare communication environment among the purchasing manager, device supplier and health users under the control of smart contracts for data traceability and access authentication.}  In \cite{111}, a secure cloud-assisted e-health system based on blockchain is proposed to protect the operation of outsourcing EHRs among medical users. This can be done by an Ethereum blockchain platform to manage user transactions without requiring any trusted entity. Meanwhile, the study in \cite{112} uses blockchain for building a healthcare remedy evaluation system on cloud. In this context, the decentralized replication of blockchain can improve the information credibility and the quality of crowdsourcing systems. Recently, a healthcare project is implemented in Peru using the BCoT platform for purchase management in private health sector \cite{113}. In this project, blockchain hosted inside the Amazon cloud is used to organize a secure communication network of purchasing manager, the supplier and the transporter. Sensor data will be authorized by smart contracts available on blockchain to avoid data alternation risks. 

Meanwhile, the authors in \cite{114} highlight the efficiency of BCoT models in health monitoring services that are enabled by IoT devices, cloud computing and blockchain. Cloud connectivity offers substantial medical computing services, such as storage and intelligent computation. 
In a recent work \cite{115}, we also introduce a conceptual BCoT framework for health diagnosis and monitoring. In particular, we integrate the data management system with a data sharing framework in a mobile blockchain network. Data is ensured security through an access control layer managed by smart contracts for access verification and data integrity. 

\textit{1.4) Lessons learned}

The main lessons acquired from the review of BCoT applications in healthcare are highlighted in the following.
\begin{itemize}
	\item \textcolor{black}{BCoT can achieve secure data sharing on cloud IoT-enabled healthcare networks where blockchain and cloud play a significant role in controlling user access and implementing data sharing.} Smart contracts available on blockchain are particularly useful to track automatically transactions and implement access verification which ensures reliability and security for untrusted healthcare environments. 
	\item \textcolor{black}{The integration of blockchain in cloud computing significantly improves security for cloud healthcare storage services. Cloud storage acts as a peer in the P2P network under the management of blockchain. In this context, original health data can be encrypted and kept in cloud storage,} while metadata (e.g., hash values) of such data records is stored in blockchain, which enables data traceability and detects easily data modification threats on cloud.
	\item BCoT can offer innovative healthcare services with high security and efficiency. BCoT has the potentials to improve the quality of medical services such as health monitoring, patient diagnosis or healthcare remedy evaluation. 
\end{itemize}

\subsubsection{Smart City}

With recent advances of cloud computing and IoT technology, smart city has been emerged as a new paradigm to dynamically exploit the resources in cities from ubiquitous devices and provide a wide range of services for  citizens. Smart cities involve a variety of components, including ubiquitous IoT devices, heterogeneous networks, large-scale data storage, and powerful processing centres such as cloud computing for service provisions. Despite the potential vision of smart cities, how to provide smart city services with high efficiency and security remains a challenging problem. In this scenario, BCoT can be a promising candidate to empower smart city services by using attractive technical features of cloud computing and blockchain. A number of recently proposed solutions suggest to adopt BCoT architectures to enable ubiquitous connectivity between citizens and industrial applications for smart cities. We summarize recent research efforts in the adoption of BCoT in smart cities via two main services: security services for smart city and smart home services.

\textit{2.1) Security services for smart city}

Due to the ubiquitous nature of data-based services, smart city architectures remain security bottlenecks such as privacy, integrity, trust, and so on \cite{116}. {The BCoT model with high security capabilities enabled by blockchain promises to help overcome such challenges as well as offer new smart city services. In such contexts, blockchain plays an important role in providing high-quality security services for smart city scenarios. In fact, blockchain is possible to provide five main cryptographic primitives, including integrity, authenticity, confidentiality and non-repudiation, by building decentralized security architectures for smart cities.}

The work \cite{117} introduces a decentralized big data integrity auditing framework in cloud environments for smart cities. The proposed architecture consists of two main entities: data owners and cloud service providers (CSPs). An innovative blockchain instantiation named the data auditing blockchain (DAB) is proposed to investigate auditing requests between users and CSPs to verify data integrity. Also, the study \cite{118} considers an authorization and delegation architecture for the cloud IoT based on blockchain in smart city projects. The mechanism is conducted in a single smart contract which enables access control functionalities to ensure trustfulness and auditing for operations between IoT devices, cloud and data users. 

Furthermore, blockchain is adopted in \cite{119} to build an IoT-based smart city architecture which includes three main layers: smart block, P2P network and cloud. Since blockchain is inefficiently for a large number of network nodes, e.g., IoT devices in smart city, the proposed scheme suggests a lightweight blockchain with low computation and resource demands. All communications between IoT devices, cloud storage and P2P nodes are tagged as transactions which are recorded and stored securely on blockchain in a tamper-proof manner. A blockchain-based infrastructure is also considered in \cite{120} to support secure smart contract services for sharing economy in smart cities. Multimedia payload from IoT devices is offloaded and stored securely in distributed IPFS-based cloud repositories as immutable ledgers. 

\textit{2.2) Smart home services}

In the context with smart cities, home automation gives shape to a smart home which is the main feature of a smart city. A smart home is a network of IoT devices configured with automated devices, intelligent sensors and detectors, which will collect information from the environment to be processed by a control server such as a computer or a cloud computing platform. Despite many potentials to benefit citizens, smart homes remain unsolved issues in terms of security, threats, attacks, and data privacy. BCoT run by blockchain that owns distributed, secure and private properties would be the promising solution to these security issues \cite{121}. {In this context, blockchain is used to build a decentralized data integrity architecture for ensuring high stability and reliability of the whole system without the requirement of third-party auditors.}

The work in \cite{122} propose a smart home architecture using a BCoT model which includes three main tiers: cloud storage, overlay (blockchain-based P2P network), and smart home. Resourceful devices act as blockchain miners to handle transactions within the smart home and ensure security objectives in terms of confidentiality, integrity, and availability. The storage of data within a smart home is implemented by cloud computing under the management of blockchain miners through a transaction authentication process which enables high security for smart home operations. 

In \cite{123}, a secure and efficient IoT smart home architecture is proposed by taking advantage of cloud computing and blockchain technology. The general structures contain four components, namely smart home layer, blockchain network, cloud computing, and service layer. Blockchain with its decentralized nature is integrated in distributed cloud storage for data usage traceability. Besides, it also serves processing services and makes the transaction copy of the collected user data from smart home.  Moreover, the study in \cite{124} also presents a conceptual access control scheme for smart home where private blockchain is used to store records of user transactions and large-size access data is stored in off-chain storage, such as cloud storage. 

\textit{2.3) Lessons learned}

The main lessons acquired from the review of BCoT applications in smart city are highlighted in the following.
\begin{itemize}
	\item BCoT can offer advanced security services for smart city applications. \textcolor{black}{Cloud computing is capable of providing powerful computing capacities to handle large data streams from ubiquitous IoT devices to offer real-time applications for citizens.} Meanwhile, with high security properties, blockchain proves its high efficiency in controlling smart city operations in a distributed and secure manner. The integration of blockchain and CoT thus transforms smart city architectures to overcome challenges in terms of security and system performance. 
	\item	As a significant component of smart city, smart home also gains benefits from the BCoT integration. BCoT can enable intelligent services, such as user monitoring, home management and access control in smart home scenarios. \textcolor{black}{ Specially, blockchain can be integrated with distributed cloud computing to make data storage and transaction processing more flexible and secure among IoT devices, home owner and external users. }
\end{itemize}
\subsubsection{	Smart Transportation}

With the rapid development of modern sensing, communicating, computing technologies, recent years have witnessed tremendous growth in intelligent transportation systems (ITS), which impose significant impacts on various aspects of our lives with smarter transport facilities and vehicles as well as better transport services. Smart transportation is regarded as a key IoT application which refers to the integrated architectures of communication technologies and vehicular services in transportation systems. One critical issue in smart transportation is security risks resulted by dynamic vehicle-to-vehicle (V2V) communication in untrusted vehicular environments and reliance on centralized network authorities. {Blockchain has the potential to help establish a secured, trusted and decentralized ITS ecosystem. The combination of cloud computing with limitless data management capabilities and blockchain with high security features is able to enhance security and quality of services in smart transportation, including vehicular communication management and secure vehicular services.} 

\textit{3.1) Vehicular Communication Management}

{In vehicular communication management, blockchain is used in information and energy interaction processes to achieve high security levels such as user authentication with contracts and data confidentiality with cryptography. During the information and energy interactions, the vehicular records are encrypted and appended into the blocks by a consensus mechanism in the blockchain network. Also, blockchain is able to create a secure peer-to-peer network to enable seamless communications among ubiquitous vehicles for service management, value exchange and collaborative trust.}
The work \cite{125} proposes a collaboration network of multiple vehicle clouds where blockchain is applied to establish a coordination scheme. Vehicles from different car manufacturers can achieve efficient interconnection through their private cloud based on a decentralized mechanism which enables service management, value exchange and collaborative trust within the vehicle-to-vehicle (V2V) communication network. Blockchain is adopted to support peer-to-peer collaboration among different clouds of vehicles with high security levels \cite{126}.

\begin{figure}
	\centering
	\includegraphics[height=8cm,width=8.8cm]{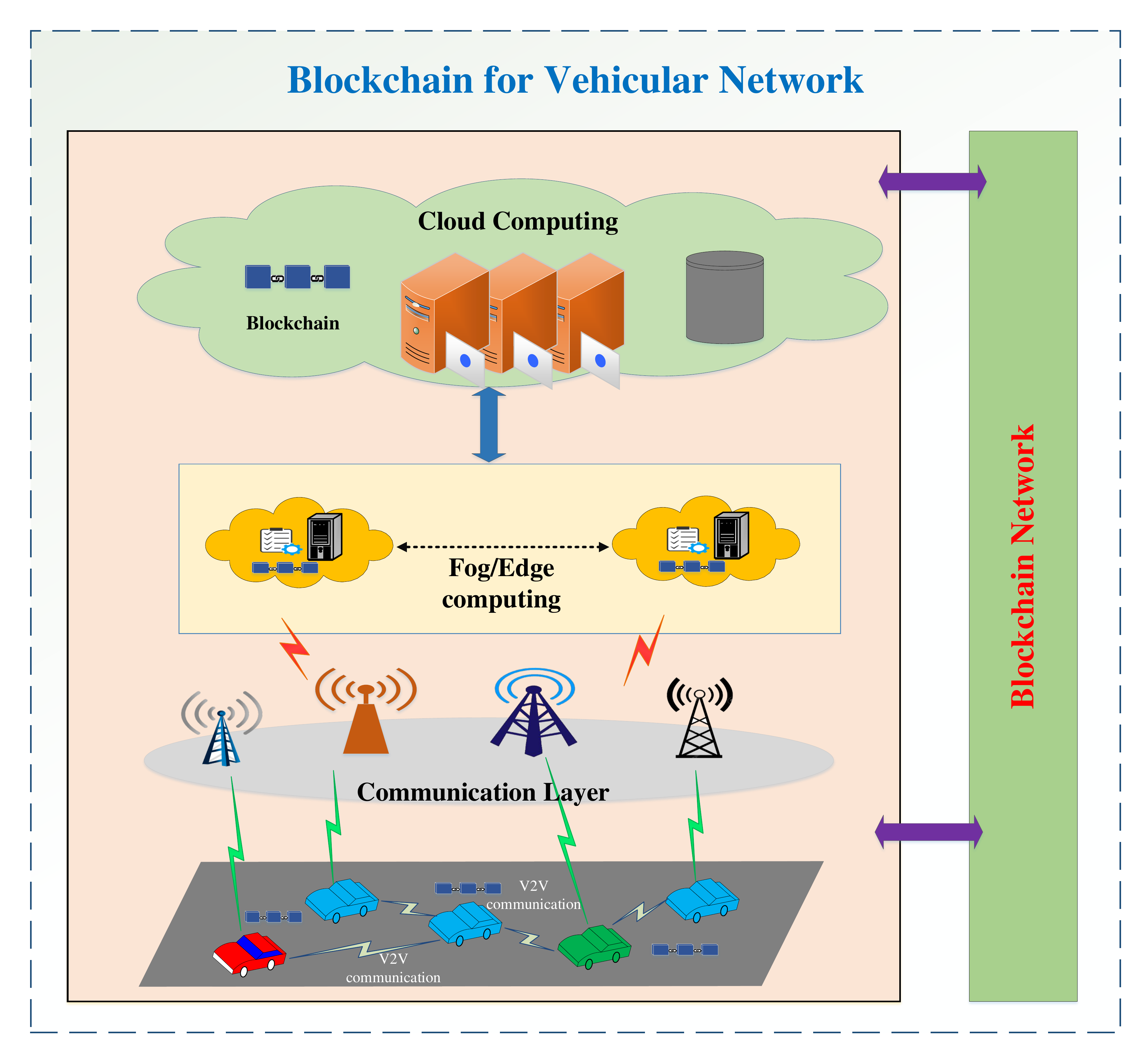}
	\caption{Blockchain and cloud for security of VANET system \cite{127}.}
	\label{vehicle}
	\vspace{-0.15in}
\end{figure}

In \cite{127}, a distributed blockchain cloud architecture is proposed to preserve privacy of vehicle drivers with on-demand and low-cost access in vehicular ad-hoc networks (VANETs). To solve issues related to limitations of storage, computation and spectrum bandwidth in VANETs, a cloud computational hierarchical architecture is proposed with three interconnected cloud platforms, consisting of vehicular cloud, road side cloud and central cloud. The joint cloud network is interconnected securely with vehicles, service providers through a P2P network run by blockchain which enables the vehicular ecosystem to be resistant to cyber-attacks and privacy bottlenecks. Another research effort \cite{128} shows a security architecture of VANETs based on blockchain and edge-cloud computing. The architecture consists of three main layers, namely perception layer with a network of vehicles, edge computing layer and service layer as illustrated in Fig.~\ref{vehicle}. Here, the service layer is established by the integration of cloud computing and blockchain to build a secure decentralized vehicular management architecture. Therefore, cloud-based huge data storage and blockchain-based data privacy are ensured for efficient and secure vehicular communication \cite{129}. 

\textit{3.2) Secure vehicular services}

{Blockchain with its decentralization and traceability also facilitates secure vehicular IoT services, from task scheduling, data carpooling, insurance management to vehicular report and trust control services. }
In \cite{130}, a BCoT architecture is developed to build a vehicular ecosystem where smart vehicles, equipment manufacturers and cloud storage providers can communicate together. The system operates under the management of a public blockchain which is possible to protect the privacy of users and to increase the security of the vehicular network. Two applications including wireless remote software updates and dynamic vehicle insurance fees are considered to demonstrate the efficiency of the architecture. 

The authors in \cite{131} consider the privacy issues of carpooling services. To achieve high security and privacy of the service, they propose an efficient and privacy-preserved scheme by using blockchain-enabled vehicular fog computing. Particularly, carpooling data is encrypted and kept at the cloud server, while its hash value is stored on the private blockchain, which enables data traceability and reliability. The work \cite{132} uses the BCoT model to build a mechanism of task scheduling in a vehicular cloud computing environment. An autonomous vehicular cloud (AVC) ecosystem is established where non-repudiation of task execution between task senders and task runners (vehicles) is guaranteed by secure transaction management of blockchain. Meanwhile, the study \cite{133} presents a fine-grained transportation prototype for insurance services enabled by the blockchain and CoT. The system consists of two main parts, namely an IoT based data collection and processing scheme for driving behaviour analytics and a collaborative blockchain network of Ethereum and Hyperledger Fabric platform for vehicular operation management. 

Furthermore, security for vehicular IoT services has become a critical challenge. The work \cite{134} proposes a blockchain-based security framework to support vehicular IoT services, e.g., real-time cloud-based video report and trust management on vehicular messages. A software-defined networking architecture is incorporated into the VANET to enable global information collection and network management. A blockchain platform is employed to build a semi-decentralized trust management architecture in which encrypted videos or messages are uploaded to the cloud for large storage while trusted traffic information is stored securely in the blockchain. 

\textit{3.3) Lessons learned}

The main lessons acquired from the review of BCoT applications in smart transportation are highlighted in the following.
\begin{itemize}
	\item BCoT paradigms can offer advanced solutions by combining cloud computing and blockchain in vehicular networks to achieve efficient and secure vehicular communication. Blockchain can create secure peer-to-peer network environments to enable limitless communications among ubiquitous vehicles for service management, value exchange and collaborative trust.
	\item CoT also enable a new set of vehicular services with better efficiency and security. \textcolor{black}{ The combination of autonomous vehicular cloud and blockchain opens up new opportunities to facilitate vehicular IoT services, from task scheduling, data carpooling, insurance management to vehicular report and trust control services, which promise to transform intelligent transportation systems. }
\end{itemize}

\subsubsection{Smart Industry}

Blockchain has emerged as an enabling technology enabled by the decentralized P2P network structure to drive smart industries, and the convergence of CoT and blockchain as a BCoT paradigm promises to empower industry ecosystems with enhanced security and improved industrial operation efficiency. There is a vast body of research works in the combination of BCoT in smart industry and we can categorize them into three areas: smart manufacturing, smart energy, and smart supply chain. 

\textit{4.1) Smart manufacturing}

\begin{figure*}
	\centering
	\includegraphics[height=10cm, width=16cm]{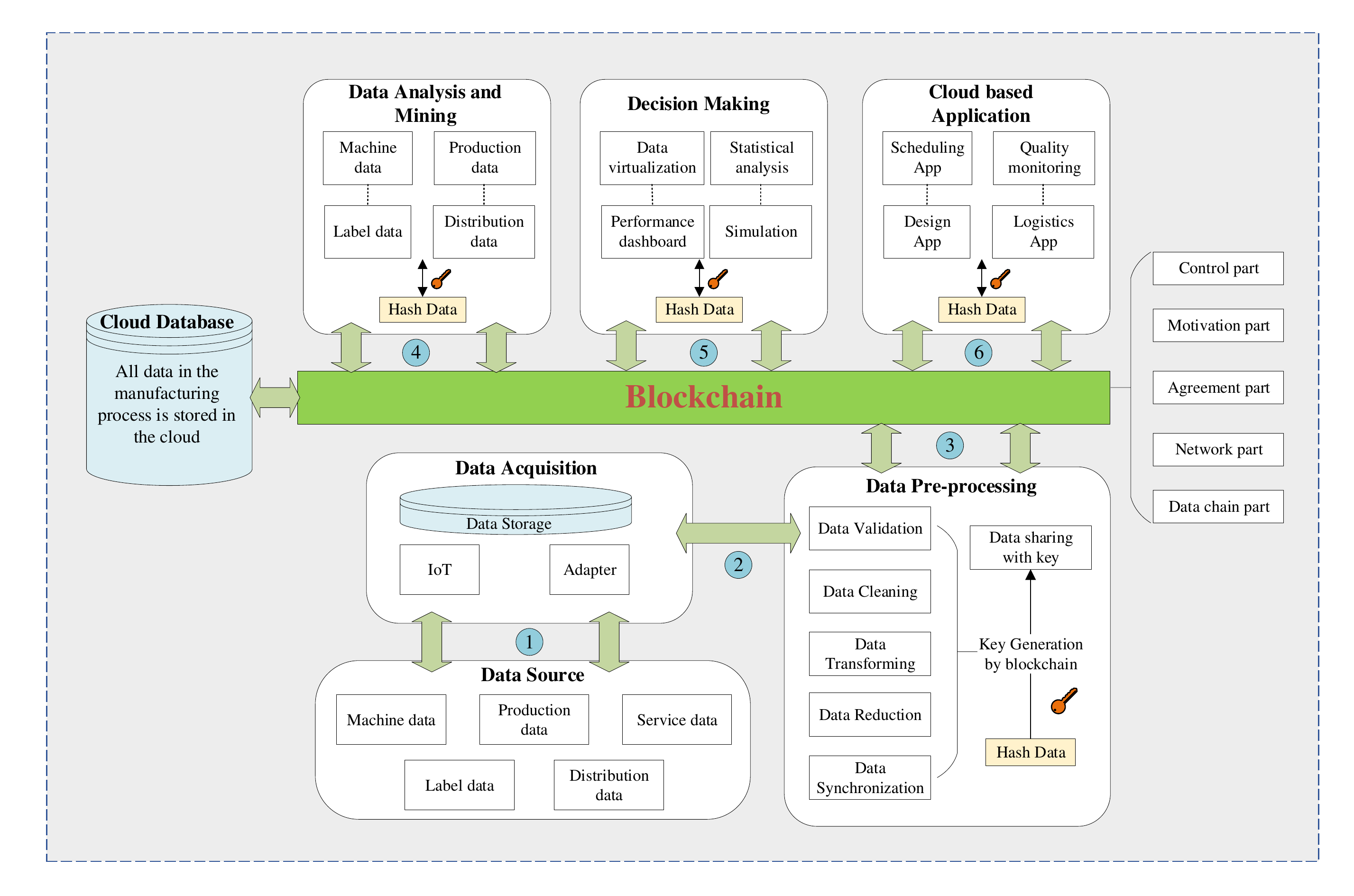}
	\caption{The blockchain cloud manufacturing system \cite{136}.}
	\label{industry}
	\vspace{-0.17in}
\end{figure*}

Smart manufacturing is a broad category of manufacturing that employs cloud manufacturing, IoT enabled technologies and service-oriented manufacturing, which benefit the manufacturing industry. However, all existing paradigms still face the main problem related to centralized industrial network and third part-based authority. In a nutshell, centralized manufacturing architectures exist limitations with low flexibility, efficiency, and security. The use of BCoT in manufacturing systems can be a promising solution to overcome such critical challenges with the support of cloud computing and blockchain as shown in Fig.~\ref{industry}. BCoT is possible to enhance and optimize manufacturing processes and reduce operation costs \cite{135}. {Particularly, blockchain can improve the security of smart manufacturing process, by offering efficient security services for trust and privacy establishment among different manufacturing enterprises. }

In \cite{136}, a distributed P2P network architecture named BCmfg is proposed with five key layers, namely resource layer, perception layer, manufacturing layer, infrastructure layer and application layer. Blockchain is integrated in the manufacturing industry to facilitate cloud manufacturing and establish a new trustable platform as blockchain cloud manufacturing. Service providers and customers can share data and information over the cloud blockchain network which helps improve the security of industry system. Smart contracts act as agreements between the end users and the service providers to provide on-demand manufacturing services. 

The work \cite{137} introduces a decentralized framework called BPIIoT for industrial IoT based on blockchain. The BPIIoT platform is regarded as a technical enabler for cloud-based manufacturing (CBM) which offers ubiquitous and on-demand network access to manufacturing resources. Blockchain is deployed to establish a peer-to-peer network for BPIIoT in which smart contracts are deployed. Here, the smart contracts work as agreements between the service consumers and the manufacturing resources to provide on-demand manufacturing services. 

\textit{4.2) Smart supply chain}

In addition to manufacturing applications, BCoT is also beneficial to industrial supply chain that is the key component in the vertical smart industry ecosystem. {Indeed, BCoT with high decentralized and immutable natures of blockchain can ensure faster and more secure corporation between companies and manufacturers in supply chain and logistic activities. \cite{138}}. Besides, it also enables secure support planning, scheduling, and monitoring supply chain operations. 

For example, the work in \cite{139} investigates the use of blockchain for transaction processing to provide different cloud-blockchain platforms for supply chain applications. Blockchain systems are divided into three categories, including private versus public, centralized versus decentralized and peer-to-peer cloud-based systems. Cloud computing can offer a number of flexible solutions for blockchain-based supply chain, from a single repository for the blockchain to multiple peer-to-peer capabilities with broad accessibility. Blockchain ensures trust among companies and businesses during supply chain operations via consensus mechanisms and smart contracts. 

\textit{4.3) Smart energy}

With the increasing demands of energy usage to support industrial and manufacturing operations, smart energy continues to play an integral part in industry ecosystems. The overall purpose of the energy system is to provide energy services to customers and companies, in a sustainable, reliable, and cost efficient manner. Information and communication technology (ICT) will be an enabler in the transition of electricity, gas and heating grids into the smart energy system. { In such a context, BCoT empowered by immutable blockchain has emerged as a promising technique to improve the security and privacy of energy exchange and transmission. }

The work \cite{140} considers the potential of cloud-blockchain technologies for decentralized operations in energy internet environments. Centralized energy management systems (EMS) tend to be inefficient to work well with a large quantity of prosumers and thus, a decentralized architecture based on blockchain is necessary to achieve high quality of services for the decentralized institutions of various energy entities. Blockchain can integrate with cloud computing to offer effective methods for information sharing and model updating in cloud-based EMS platforms. Specially, cloud computing operations for energy management are optimized and ensured high security with a decentralized verification mechanism which is made by blockchain-based consensus among energy users. 

In \cite{141}, a blockchain-based architecture is proposed to manage the operation of crowdsourced energy systems, enabling P2P energy trading at the distribution level, where ubiquitous distribution-level asset owners can trade with each other. The platform is implemented by the IBM Hyperledger Fabric network deployed in cloud to offer blockchain services. Moreover, smart contracts are used to run the pricing mechanism and control energy trading transactions and crowdsources. The work in \cite{142} also implement an intelligent energy-aware resource management in cloud datacentre. With the support of blockchain, the enery management scheme does not require any scheduler, reducing extra energy cost and increasing the robustness of DCs. Smart contracts are also employed and stored in each datacentre to verify transactions from request migration to the DC.

\textit{4.4) Lessons learned}

The main lessons acquired from the review of BCoT applications in smart industry are highlighted in the following.

\begin{itemize}
	\item BCoT demonstrates its potentials in the improvement of smart industry for better efficient manufacturing, lower operational costs and minimum management efforts \textcolor{black}{through the use of controlling capabilities of blockchain and service support of cloud computing. }
	\item BCoT with blockchain as a middle communication layer can enable faster and more efficient corporation between companies, manufacturers and users in supply chain and logistic activities. \textcolor{black}{Security and information privacy during supply chain operations can also be ensured by consensus mechanisms over the peer-to-peer network enabled by cloud-blockchain integration. }
	\item BCoT architectures can empower energy systems which are regarded as a key component of smart industry. Blockchain has potentials to improve security and privacy of energy exchange and transmission, while cloud computing offers storage and management services as well as supports blockchain in achieving decentralized energy operations. 
	
\end{itemize}
\subsubsection{Other BCoT Applications}
The application of BCoT paradigms has been investigated in other scenarios, including smart cloud services, smart resource management and smart education.  

\textit{5.1) Smart cloud services}

Cloud computing offers a diverse range of outsourcing services, including storage and computation to serve individuals and enterprises. Basically, outsourcing services usually include online payment and security issues. However, most traditional service solutions have to rely on a trusted third-party to realize fairness to complete payments. Therefore, the realization of secure and fair payment of outsourcing services is of paramount importance for cloud-based applications.  {In this regard, blockchain has emerged as a strong candidate to solve security issues of cloud services and simplify the cloud service management thanks to its traceable and immutable nature.} The works \cite{143}, \cite{144} introduce a blockchain based fair payment architecture for outsourcing services in cloud computing. The proposed system ensures to provide soundness and robust fairness capabilities by using a service management protocol run by blockchain. Fair payment can be achieved between users and outsourcing service providers on clouds through transactions which are stored and verified by blockchain without the involvement of any third party. 

During the process of outsourcing data on clouds, when a user wants to delete the outsourced data, he sends a deletion command to the cloud server so that the server delete the data. However, the cloud server is semi-trusted and it may not delete the requested data honestly due to financial incentives. To solve this issue, the study in \cite{145} presents a new publicly verifiable data deletion scheme for cloud computing enabled by blockchain, which not only supports public verification on deletion requests but also eliminates totally the need of trusted third parties. Blockchain offers fairness verification services which enable all users can authorize transactions for data deletion requests and control malicious behaviours on their cloud data with equal verification rights. This blockchain-base scheme helps reduce the dependence on cloud servers in user data management and makes the deletion operation much more transparent. 

Meanwhile, the authors in \cite{146} propose a building information modeling system model called bcBIM to address information security challenges in mobile cloud environments. Specially, blockchain is employed to facilitate BIM data audit for historical modifications of big data sharing. BIM data integrity and provenance are guaranteed by integrating blockchain in BIM database, and system management can be achieved by BIM cloud. The BIM model based on cloud blockchain promises to foster industrial applications, such as engineering machines and construction robots. For data record management on clouds, \cite{147} and \cite{148} use blockchain to build decentralized cloud storage ecosystems for security improvements. Blockchain-based distributed storage is different from traditional cloud storage services because it utilizes the disk space of a network of computers and storage facilities to decentralize the database, which ensures that any data owners can verify and check data integrity via the P2P network on blockchain. This advanced storage concept also improves trustworthiness and data availability on cloud during data long-term preservation. 

\textit{5.2) Smart resource management}

Computing resource management for CoT in blockchain network is also attracting increasing attention. { Many approaches have been proposed to enhance computation resources and security services for BCoT applications in various types of tasks such as real-time processing, resource-intensive applications, and consensus process with the help of immutable blockchain and smart contracts for monitoring resource usage and data authenticity in cloud computing.}

The work \cite{149} introduces an optimal computing resource allocation based on an auction scheme for edge-cloud-enabled IoT in the blockchain network. A pure P2P computing resource trading system on clouds is built to establish computing resource trading between resource sellers and buyers. Meanwhile, a lightweight infrastructure of the PoW-based blockchains are proposed \cite{150} so that the workload of mining process is offloaded to the cloud/fog for computation. The computation resource management in the blockchain consensus process is formulated as a two-stage Stackelberg game, where the profit of the cloud/fog providers (CFPs) and the utilities of individual miners are jointly optimized.  

Further, the authors in \cite{151} introduce a resource management system called Saranyu which adopts smart contracts to control tenant and service accounts as well as monitor resource usage in a cloud computing data center. Saranyu is capable of offering four different services: identity management, authentication, authorization on service resource exploitations, and charging. 

\textit{5.3) Smart education}

The application of BCoT to the education domain is still in its early stages. Only a small number of educational institutions have started to utilize BCoT technology. Most of current solutions use BCoT for the purpose of validating and securely sharing academic certificates and personal information of students as well as learning database of educational institutions \cite{152}. {Immutable blockchain ledgers and cloud computing can be combined to develop secure and trusted educational environments for promoting educational collaboration.}

For example, the study in \cite{153} proposes an online identity verification system using blockchain to implement cloud educational collaboration. To achieve time authentication on transactions, blockchain is adopted to provide proofs of data content originality, which also maintains system integrity. Some case studies were conducted at Chuo University in Japan to verify the efficiency of the proposal. Additionally, blockchain ledgers and cloud computing are considered to support computer science education in \cite{154}. A new decentralized P2P-cloud model is also proposed using Bitcoin and Torrent models to build proof-of-concept platforms to support service providers in education.  

\textit{5.4) Lessons learned}

 The main lessons acquired from the review of the above BCoT applications are highlighted in the following.
\begin{itemize}
	\item The BCoT architecture has great potentials to transform cloud-based services with better efficiency and security levels. Blockchain can be involved in cloud management processes by autonomous consensus mechanisms and control capabilities of smart contracts, which ensure data integrity, data availability and trustfulness. \textcolor{black}{Importantly, blockchain helps to build peer-to-peer architectures and integrate them in cloud platforms to enable decentralized cloud services with advantages over conventional cloud ecosystems in terms of no single points of failure for better system robustness and low communication overheads.} 
	\item BCoT also proves its benefits to network resource management with a wide range of services on real-time processing, resource-intensive applications, blockchain mining, and consensus process. With the support of BCoT and smart contracts available on blockchain, resource management systems can achieve high transparency, trustfulness and ensure robust access control in collaborative networks of resource providers and customers. 
	\item Besides, BCoT can be useful to education management ecosystems. Blockchain ledgers are able to provide secure and trusted educational environments for promoting educational collaboration.
\end{itemize}

In summary, we list BCoT applications in the taxonomy Table III and Table IV to summarize the contributions and limitations of each reference work. 

\begin{table*}
	\centering
	\caption{Taxonomy of BCoT applications.  }
	\label{table}
	\begin{tabular}{|p{1.5cm}|c|P{1.9cm}|P{1.3cm}|P{5cm}|P{5cm}|}
		\hline
		\centering \textbf{Category}& 	
		\textbf{Ref.} &	
		\centering \textbf{Use case}&	
		\centering \textbf{Blockchain platform}& 
		\textbf{Main contributions}	&  
		\textbf{Limitations} 	
		\\
		\hline
		\multirow{30}{1.5cm}{Smart healthcare} & 
		\cite{99}& 	EMRs sharing& Ethereum& 	An EMRs sharing scheme to ensure data privacy on cloud.	& The real prototype is not implemented between cloud, blockchain and medical users.\\
		\cline{2-6}&
		\cite{100}&	Health data sharing&	Hyperledger Fabric &	A user-centric health data sharing solution for cloud healthcare with a focus on scalability and data integrity evaluation.&	Security issues on healthcare IoT devices, e.g., malicious attacks, are not considered. \\
		\cline{2-6}&
		\cite{101}&	Data sharing, access control&	Ethereum &	A data sharing with access control in decentralized loud storage. &	Issues on data confidentiality and access control latency are not discussed in detail.\\
		\cline{2-6}&
		\cite{102}&	Data sharing &	-&	A lightweight data sharing scheme in cloud blockchain.&	The real prototype is not investigated between cloud, blockchain and medical users. \\
		\cline{2-6}&
		\cite{103}&	Trust-less data sharing	&-&	An access control mechanism to track data access behaviours of cloud providers.&	Implementation results on access control efficiency is not investigated. \\
		\cline{2-6}&
		\cite{105}&	E-health data sharing&	Ethereum 	&A mobile cloud blockchain platform for e-health sharing with an access control design.&	Data confidentiality and scalability are not considered in detail. \\
		\cline{2-6}&
		\cite{106}&	healthcare data management &	Ethereum &	A privacy-preserved platform for data storage in cloud.	&Comparisons between smart contract-based scheme and conventional schemes have not been done. \\
		\cline{2-6}&
		\cite{107}&	Cloud data storage 	&-	&A blockchain-based cloud storage scheme for data integrity and traceability.	&Smart contract implementation on data storage has not been considered. \\
		\cline{2-6}& 
		\cite{108}&	Cloud data storage	&-&	A EMRs storage scheme on blockchain-based cloud.&	Investigations on blockchain prototype has not been done.\\
		\cline{2-6}& 
		\cite{109}&	Security for EMRs storage	&-&	A security scheme for EMRs storage.&	Real experiments on the proposed security scheme has not been done.\\
		\cline{2-6}& 
		\cite{111}&	Secure healthcare service&	Ethereum &	A secure cloud-assisted e-health system.	&Smart contract design for service management has not been considered.\\
		\cline{2-6}& 
		\cite{112}	&Healthcare remedy service&	-	&A healthcare remedy evaluation system.&	Performance for blockchain implementation on cloud has not been done.\\
		\cline{2-6}& 
		\cite{113}& 	Medical supply chain	& Hyperledger& 	Purchase management in private health sector.& 	Data privacy has not been considered. \\
		\cline{2-6}& 
		\cite{114}& 	Health monitoring services& 	-& 	A concept of health monitoring services using BCoT approach.& 	Performance evaluation on the proposed scheme has not been done.\\
		\cline{2-6}& 
		\cite{115}& 	Health diagnosis and assessment	& Ethereum& 	A conceptual framework on health assessment and monitoring using cloud blockchain& 	System scalability and communication costs have not been considered.  \\
		\cline{2-6}
		
		\hline
		\multirow{13}{*}{Smart city} & 
		\cite{117}& 	Data auditing 	& -	& A decentralized data auditing framework on cloud for smart cities.& 	Smart contract design, experiment on security evaluation have not been done.\\
		\cline{2-6}&
		\cite{118}&	Service authorization and delegation &	Ethereum&	An authorization and delegation scheme for BCoT-enabled smart city. 	&Privacy is not taken into consideration. \\
		\cline{2-6}&
		\cite{119}&	Secure smart city architecture	&Ethereum&	A BCoT smart city platform for high security.&	Access control for cloud storage has not been considered. \\
		\cline{2-6}&
		\cite{120}&	Sharing economy services&	Ethereum and Hyperledger&	A blockchain-based infrastructure for secure sharing economy services. &	Data privacy has not been analysed. \\
		\cline{2-6}&
		\cite{122}&	Smart home services	&-&	A BCoT architecture for security and privacy in smart homes.&	Blockchain implementation has not been done.\\ 
		\cline{2-6}&
		\cite{123}&	Smart home services	&-&	A smart home architecture for security services.&	Real blockchain implementation has not been investigated. \\
		\cline{2-6}
		\hline
	\end{tabular}
\end{table*}

\begin{table*}
	\centering
	\caption{Taxonomy of BCoT applications (continued).  }
	\label{table}
	\begin{tabular}{|P{1.5cm}|c|P{1.9cm}|P{1.3cm}|P{5cm}|P{5cm}|}
		\hline
		\centering \textbf{Category}& 	
		\textbf{Ref.} &	
		\centering \textbf{Use case}&	
		\centering \textbf{Blockchain platform}& 
		\textbf{Main contributions}	&  
		\textbf{Limitations} 	
		\\
		\hline
		\multirow{20}{1.5cm}{Smart transportation} & 
		\cite{125} &Vehicle collaboration&	-	&A joint cloud collaboration scheme between vehicle clouds based on blockchain&	Blockchain implementation has not been done. \\
		\cline{2-6}&
		\cite{126}&	Information and energy interactions	&- &	An electric vehicles cloud and edge computing network paradigm for secure vehicular communication with blockchain.&	Network performance has not been evaluated in experiments. \\
		\cline{2-6}&
		
		\cite{127}&	Secure vehicular communication&	-&	A distributed vehicular network based on cloud blockchain for data privacy.&	Blockchain implementation for the proposed approach has not been done.\\
		\cline{2-6}&
		\cite{128}&	Secure vehicular communication	&-&	A multi-layer decentralized VANET architecture for secure vehicular communication. &	Implementation to investigate the system efficiency is lacked. \\
		\cline{2-6}&
		\cite{130}&	Secure vehicle services	&-&	A BCoT platform for secure vehicular services, e.g., remote software updates and vehicle insurance fees.&	Blockchain and smart contract implementations have not been done.\\
		\cline{2-6}&
		\cite{131}&	Carpooling services	&-&	A privacy-preserving carpooling scheme using blockchain with cloud-fog computing. &	Scalability of the proposed scheme has not been verified. \\
		\cline{2-6}&
		\cite{132}&	Vehicular task scheduling &	Ethereum &	A strategy for task scheduling in autonomous vehicular cloud system.&	The performance of the proposed framework has not been simulated. \\
		\cline{2-6}&
		\cite{133}	&Fine-grained transportation insurance&	Ethereum and Hyperledger&	A fine-grained transportation scheme for insurance services on cloud blockchain. &	Privacy issues in vehicular transactions is not taken into consideration. \\
		\cline{2-6}&
		\cite{134}	&Vehicular IoT services	&-	&A blockchain-based security framework for vehicular IoT services.&	Scalability of the proposed scheme and access control have not been verified.\\
		\cline{2-6}
		\hline

		\multirow{14}{*}{Smart industry} & 
		\cite{136}&	Smart manufacturing &	Ethereum&	A blockchain cloud manufacturing system for secure manufacturing industry. 	&Access control to the data storage in cloud database has not been considered.\\
		\cline{2-6}&
		\cite{137}&	Manufacturing platform&	Ethereum&	A decentralized framework for manufacturing applications with cloud blockchain. &	The performance of the proposed framework has not been simulated.\\
		\cline{2-6}&
		\cite{139}&	Supply chain&	-&	A conceptual cloud blockchain framework for supply chain. &BCoT implementation for supply chain scenarios has not been done. \\
		\cline{2-6}&
		\cite{140}	&Smart energy&	-&	A cloud-blockchain scheme for decentralized operations in energy internet. &	Only conceptual analysis is provided and simulation to evaluate the proposal is lacked.\\
		\cline{2-6}&
		\cite{141}	&Smart grids&	Hyperledger Fabric&	A crowdsourced energy system for energy trading on cloud blockchain.&	Scalability of the cloud-blockchain based solution in smart grids has not been considered. \\
		\cline{2-6}&
		\cite{142}&	Smart energy&	Ethereum&	An intelligent energy aware resource management in cloud datacentre (DC) using blockchain. &	Mining cost, privacy of energy data in cloud data centre have not been considered.  \\
		\cline{2-6}
		\hline
		
		\multirow{14}{*}{Cloud services} & 
		
		\cite{143}&	Smart payment&	Ethereum&	A blockchain based fair payment architecture named BCPay for outsourcing services.&	Data privacy has not been investigated. \\
		\cline{2-6}&
		\cite{145}&	Data delection&	-&	A new publicly verifiable data deletion scheme on cloud using blockchain.&	The feasibility of the proposed model has not been investigated on real world cloud platforms. \\
		\cline{2-6}&
		\cite{149}&	Resource management &	-&	An optimal computing resource allocation scheme on cloud blockchain in IoT.&	Smart contracts for access control among resource users have not been investigated.  \\
		\cline{2-6}&
		\cite{151}	&Cloud service management&	Ethereum&	A service management system based on smart contracts for cloud resource provisions. &	Data privacy on service exchanges has not been investigated.\\
		\cline{2-6}&
		\cite{153}&	Smart education	&-&	An online identity verification system based on blockchain for educational collaboration.&	Real implementation results have not been reported to verify the proposal. \\
		\cline{2-6}
		\hline
	\end{tabular}
\end{table*}

\subsection{BCoT Platforms and Services }
The integration of blockchain and CoT can lead to develop unprecedented architectures to enable smart services across IoT domains. In this section, we review the latest research efforts to integrated BCoT models with cloud blockchain storage platforms and cloud services for BCoT applications.
\subsubsection{Cloud Blockchain Platforms}
The data storage of traditional CoT applications have mainly relied on cloud computing which is a completely central environment. This centralized storage architecture shows a number of critical limitations, such as the lack of user control on IoT data as well as security and privacy concerns. Further, centralized cloud storage service providers charge a significant fee for their services. For example, Amazon cloud charges \$23 a month for the storage service they provide, which may pose a burden on small cloud IoT projects. On the other hand, conventional blockchain seems to be very expensive for storing large amounts of IoT data on chain. In fact, blockchain platforms like Bitcoin are restricted data storage only one megabyte. To overcome such challenges, decentralized storage based on cloud blockchain would be a promising solution which offer highly flexible, secure, trustful and super cheap storage services for BCoT applications \cite{101}, \cite{105}. 

\begin{table*}[ht]
	\centering
	\caption{Decentralized storage platforms based on cloud blockchain. }
	\label{table}
	
	\setlength{\tabcolsep}{5pt}
	\begin{tabular}{|c|p{5.6cm}|c|c|c|c|c|}
		\hline
		\textbf{Platforms}&  \centering 
		\textbf{Key features}&	
		\textbf{Cloud support}& 
		\centering \textbf{Latest version}&
		\centering \textbf{Last update}& 
		\textbf{Ready to use?}&
		\textbf{Open source?} 		
		\\
		\hline
		IPFS &	Data file is hashed cryptographically for immutability. 	&Yes&	v0.4.21 & May 2019&	Yes &	Yes \cite{176}
		\\
		\hline
		Storj &	End-to-end encryption security is provided. &	Yes &	v04	&Apr. 2019	&Yes &	Yes \cite{177}
		\\
		\hline
		Filecoin&	End-to-end encryption security is provided. Users can stored files with preferences based on cost budgets, redundancy, and file retrieval speeds&	Yes&	v 0.2.2	&May 2019	&Yes&	Yes \cite{178}
		\\
		\hline
		Sia&	Stored files are encrypted. Storage is supper cheap with \$2/ terabyte. &	Yes	&v1.3.3&	Aug. 2018&	Yes	&Yes \cite{179}
		\\
		\hline
		Swarm&Users can use local HTTP proxy API to interact with Swarm. Ethereum support is provided. &-&	v 0.4.3	&Jun. 2019&	Yes&	Yes \cite{180}
		\\
		\hline
		Maidsafe	&Data file is uploaded to a safe network and is fully encrypted for privacy. &	-&	v4.18.2	&2018	&Yes&	Yes \cite{181}
		\\
		\hline
		BigchainDB&
		It combines the key benefits of distributed databases and traditional blockchains.&	Yes	&v2.0&	2018&	Yes	&Yes \cite{182}
		
		\\
		\hline
		Datum&	Users can offload data to decentralized nodes via mobile application with smart contracts.  	&-&	v0.1.33&	2018&	Yes&	Yes \cite{183}
		\\
		\hline

	\end{tabular}
\end{table*}
In this regard, we survey the most popular decentralized storage platforms and summarize them in Table V. Key information of these platforms is also highlighted, and the open sources of software for ready usage are also released. With these advanced storage solutions, the BCoT applications do not rely on a central service provider, allowing users to store IoT data to a distributed set of storage nodes, e.g., computers, based on the peer-to-peer network on blockchain.  In fact, many of these systems have proven their efficiency in IoT scenarios. For example, the decentralized IPFS \cite{176} and Storj \cite{177} storage platforms are applied in IoT systems on cloud blockchain \cite{105}, \cite{147} and shows their efficiency in terms of low access latency and improved security levels, compared to traditional centralized storage solutions. Additionally, Swarm \cite{180} works as the distributed data storage platform running on Ethereum which has the potential to manage and share securely IoT data against distributed denial of service (DDoS) attacks and malicious access \cite{184}. Recently, some cloud giants have launched initiatives to integrate decentralized storage for large-scale cloud blockchain deployments, such as IPFS storage on Amazon and Microsoft Azure clouds \cite{105}, \cite{185}, for secure and efficient data storage. These interesting integrations have the potentials to disrupt both blockchain and cloud computing worlds to enable new infrastructures for future BCoT applications.

\subsubsection{BaaS Services for CoT}
In BCoT ecosystems, blockchain can be regarded as a Blockchain-as-a-Service (BaaS) which is integrated with cloud computing to offer full IT services in order to help researchers and enterprises develop, verify and deploy blockchain for cloud IoT applications. Specially, BaaS services are capable of providing foundation architecture and technical support to ensure that BCoT systems can achieve robust and efficient operations. Nowadays, there is a large number of BaaS providers on commercial markets to enable customers to adopt services without worrying about infrastructure installation and system investment, which can accelerate their BCoT deployments.

	\begin{table*}[ht]
	\centering
	\caption{BaaS platforms for BCoT applications. }
	\label{table}
	
	\setlength{\tabcolsep}{5pt}
	\begin{tabular}{|P{2cm}|p{8.8cm}|P{2.2cm}|c|c|}
		\hline
	\centering		\textbf{BaaS Services}& \centering	
		\textbf{Descriptions}& 
		\centering \textbf{Blockchain }&
		\centering \textbf{Launch Year}& 
		\textbf{Source code}		
		\\
		\hline
		Microsoft Azure Blockchain &	Microsoft Blockchain on Azure is a BaaS platform hosted on the Microsoft Azure cloud computing for creating and configuring consortium blockchain infrastructure quickly. It is now available in two tiers: Basic for cost-optimized services to test blockchain apps and Standard for running real BCoT applications.&	Ethereum, Hyperledger Fabric or R3 Corda&	2016	&\cite{186}				
		\\
		\hline
		IBM Blockchain &	IBM blockchain is an enterprise-ready blockchain application development platform. It enables businesses to develop, govern, and operate blockchain systems with seamless software and network updates on IBM cloud. Some biggest banking and commercial industries have used IBM blockchain.&	Hyperledger Fabric&	2017&	\cite{187}
		\\
		\hline
		Amazon Blockchain &	Amazon blockchain service makes it easy to setup, deploy, and manage scalable blockchain networks. It can be useful in many IoT use cases, such as manufacturing, insurance, trading, retail, and banking systems. 	&Ethereum and Hyperledger Fabric	&2018&	\cite{188}
		\\
		\hline
		Oracle Blockchain&	BaaS on Oracle cloud provides an enterprise-grade distributed ledger platform that can assist businesses to increase trust and provide agility in transactions across their business networks. Oracle BaaS can seamlessly connects with a number of popular Oracle solutions such as Oracle Supply Chain Management (SCM) Cloud and Oracle Enterprise Resource Planning (ERP) Cloud. &	Hyperledger Fabric	&2018	&\cite{189}
		\\
		\hline
		Hewlett-Packard (HP) &Blockchain	HP launched its BaaS called HPE Mission Critical Blockchain, which enables customers to execute distributed-ledger workloads in industrial environments with high security. It also guarantees massive scalability of HP-based blockchain projects to support business. & 	Ethereum &	2017&	\cite{190}
		\\
		\hline
		Alibaba Blockchain&	Alibaba BaaS is an enterprise-level PaaS (Platform as a Service) which is built on Alibaba Cloud Container Service for Kubernetes clusters. It brings benefits such as high security, ease-of-use, high stability, openness and efficient sharing services for blockchain-based applications. &	Ethereum and Hyperledger Fabric&	2017&	\cite{191}
		\\
		\hline
		Baidu Blockchain &	Baidu BaaS is a commercialized platform, to simplify Dapp development. It provides developers with services such as multi-chain and middle-tier frameworks, as well as smart contract and DApp templates on Baidu cloud computing. Its applications consists of IoT with BCoT, finance, and data sharing.&	Ethereum, Hyperledger Fabric, and Baidu XuperChain&	2018&	\cite{192}
		\\
		\hline
		Huawei Blockchain 	&Huawei BaaS is a cloud service that capitalizes on the advantages of Huawei cloud’s container and security technologies. It offers key advantages such as open, easy-to-use, flexible and efficient features as well as robust security and privacy protections. &	Hyperledger	&2018&	\cite{193}	
		\\
		\hline
		Google Blockchain&	Google BaaS is based on Ethereum platform with important features such as API integration, configurable consensus algorithms, and the ability to use a traditional SQL databases to query and report on blockchain data. &	Ethereum&	2018&	\cite{194}
		\\
		\hline
		SAP&	SAP BaaS provides an easiest and lowest-risk gateway to experimenting with distributed ledger technology. It is hosted on SAP cloud platform, enabling to prototype, test, and build blockchain applications (both private and consortium) and smart contracts. 	&MultiChain and Hyperledger Fabric&	2018&	\cite{195}
		\\
		\hline

	\end{tabular}
\end{table*}

Reviewing thoroughly the state-of-the-art BaaS platforms available on the market, in this subsection, we introduce the leading BaaS platforms which are ready to use for BCoT applications. The key technical characteristics of each platform are described briefly in Table VI. The source code for BaaS examples and templates are also available on the code sharing platform Github. In fact, many research projects have employed such BaaS platforms to develop their BCoT applications. For example, the Amazon Blockchain service \cite{188} is adopted to build an IoT healthcare system \cite{105}. In this project, the Ethereum blockchain platform hosted on Amazon cloud helps to implement a health data sharing framework on mobile clouds with high security and privacy.  Moreover, IBM cloud \cite{187} also introduces a well-developed BaaS platform for IoT users. The platform has been showcased in a vehicular network \cite{208}. In this project, the IBM IoT platform is integrated with IBM BaaS services to manage vehicle sensor data (vehicle-to-vehicle messages and vehicle monitoring data) and ensure security during data sharing within the vehicular network. Meanwhile, the BaaS platform of Oracle cloud \cite{189} has proven its great potentials through a wide range of BCoT projects, such as banking, healthcare data management, and payment industry \cite{209}. Recently, the Hewlett Packard cloud provider \cite{190} collaborates with the automotive manufacturing giant Continental to launch a blockchain-based platform for car manufacturers to share and sell vehicle data \cite{210}. This project allows customers, including vehicle drivers, car manufacturers and service providers can share securely vehicle data in untrusted vehicular networks, making mobility safer, greener, and more accessible. Although the development of BaaS platforms is still in progress, the success of such initial projects on BaaS platforms is expected to open up new opportunities for future BCoT deployments as well as disrupt global industries.

\section{\textcolor{black}{Research Challenges and Discussion }}
From the extensive review on BCoT integrations, we identify possible research challenges and open issues in the field. We also discuss some potential solutions to encourage more research efforts in this promising area.

\subsection{Research Challenges} 
We highlight five major challenges in BCoT research, namely standardization, security vulnerability, privacy leakage, intelligence, and resource management.
\subsubsection{Standardization}
Since its inception, the blockchain technology has revolutionized industries by offering new network models with its decentralized and secure natures. The arrival of this emerging technology is potential to change the current shape of CoT markets and transform industrial network architectures with advanced BCoT paradigms. Although the convergence of blockchain and CoT can bring various benefits to IoT applications, the BCoT technology has developed without standards and is limited to a few service providers. Importantly, each service provider mainly designs and offers BCoT for specific applications rather than generic schemes which can be applicable to diverse use-case domains. The lack of system standard can restrict potential collaborations between services providers and make customers feel difficult in changing providers as each provider has their own rules \cite{27}. Furthermore, non-standard heterogeneous communication protocol between different blockchain platforms and CoT systems is still a critical issue for the BCoT market. For example, three cloud blockchain projects considered in \cite{27} including Golem, SONM and iExec have different visions in terms of service provision, system configuration, and customer targets. Such a lack of standard arises from three main reasons: different service definitions, different network management concepts and different operational hypothesis. Consequently, they are unable to meet a standard service level agreement, which is very important for their project developments in a long run. 
 
\subsubsection{Security Vulnerability}

Although blockchain can bring security benefits to CoT thanks to its distributed nature, immutability, verifiability, and encryption, security issues in BCoT still remain due to the vulnerabilities of both CoT and blockchain systems. 
In CoT, there has been an increasing demand of outsourcing IoT data to clouds for storage and computation services due to the constrained resources of IoT devices. This dynamic incorporation has brought a series of new challenging security concerns such as identity and access control, authentication, system integrity \cite{62}. Further, there are a number of critical security attacks to CoT, such as eavesdropping, malicious IoT attacks, unsecured communication channels, and degradation of connection quality. Cloud services for BCoT also suffer from serious security threats, from storage and computation attacks, virtual machine (VM) migration attacks to malware injection and DoS attacks \cite{219}. 

On the other side, recent studies also have revealed inherent security weaknesses in blockchain operations which are mostly related to BCoT systems \cite{220}. A serious security bottleneck is 51\% attack which means that a group of miners controls more than 50\% of the network's mining hash rate, or computing power, which prevents new transactions from gaining confirmations and halts payments between service providers and IoT users. Seriously, attackers can exploit this vulnerability to perform attacks, for example, they can modify the ordering of transactions, hamper normal mining operations or initiate double spending attack, all of which can degrade the blockchain network \cite{220}. In addition, the security aspect of smart contract, which is regarded as core software on blockchain, is also very important since a small bug or attack can result in significant issues like privacy leakage or system logic modifications \cite{221}, \cite{222}. Some of critical security vulnerabilities can include timestamp dependence, mishandled exceptions, reentrancy attacks on smart contracts in BCoT applications.

\subsubsection{Privacy Leakage}
The privacy of IoT data in BCoT can be compromised accidentally and hence the disclosure of data is respectively beneficial for the attacker and harmful to the users. In current BCoT systems, data can be stored off-chain in cloud storage to reduce the burden on blockchain. However, this storage architecture can arise new privacy concerns. Specifically, an autonomous entity can act as a network member to honestly perform the cloud data processing, but meanwhile obtains personal information without the consent of users, which leads to serious information leakage issues. External attacks can also gain malicious access to retrieve cloud data, or even alter and modify illegally outsourced IoT records on cloud. Besides, privacy leakage on blockchain transactions is another significant problem. Although blockchain uses encryption and digital signature to preserve transactions, recent measure results \cite{228} show that a certain amount of transaction is leaked during blockchain operations and data protection of blockchain is not very robust in practice. Furthermore, criminals can leverage smart contracts for illegal purposes, facilitating the leakage of confidential information, theft of cryptographic keys.  Importantly, the privacy of BCoT users can not be ensured once they join the network. Indeed, by participating in the blockchain network, all information of users such as the addresses of sender and receiver, amount values is publicly available on the network due to the transparency of blockchain. Consequently, curious users or attacks can analyse such information and keep track of activities of participants, which can lead to leakage of information secrets such as personal data. 

\subsubsection{Intelligence}
Currently, BCoT systems are mainly used for data storage, data sharing and security services. However, there has been a lack of research attention in integrating intelligent services in BCoT applications. In fact, modern industries have increasing demands in intelligent services such as smart data analytics, smart decision making systems or automatic management tools to facilitate user service delivery. For example, a smart clinical support system based on cloud computing in healthcare can make diagnosis and treatment much easier. Further, an intelligent traffic analytic tool in cloud-based vehicular networks can help vehicle drivers to adjust their route for reducing possibilities of traffic congestion. All such intelligent services will be promising in BCoT-enabled applications to satisfy quality of user experience and enhance system efficiency. Therefore, we consider intelligence in BCoT as an important open issue where research efforts are strongly necessary.

\subsubsection{Resource Management}
In BCoT applications, to achieve sustainable profit advantage, cost reduction, and flexibility in cloud service provision, the  resource management in cloud blockchain is vitally important and needs more research efforts. In fact, resource management in cloud blockchain networks requires adaptive and robust designs to solve series of technical problems, from resource allocation, bandwidth reservation to task allocation and workload allocation. A set of issues, challenges, and future research directions on resource management in cloud-based networks is discussed in \cite{237}, for example, the optimization of cloud resource allocation to computation demands and the adaption to dynamic service usage patterns. Such issues would become more complex when integrating cloud computing in blockchain where the resource usage is divided to serve multiple purposes, including resource for user demands and resource for mining mechanisms to maintain blockchain. Therefore, there is an urgent need to seek innovative solutions to overcome challenges in terms of resource management in integrated BCoT networks.

\subsection{\textcolor{black}{Discussion}}
Different BCoT service providers should achieve a service agreement on the incorporation of blockchain and CoT. Technical details such as network settings, blockchain deployment, IoT device integration, and service payment schemes should be considered carefully. Federation of service providers can be necessary to standardise the BCoT technology. Many standardization efforts have been made with the participation of a number of organizations such as ISO, ISTIC Europe, IEEE \cite{217} to build a general functional architecture for blockchain platforms. Moreover, international BCoT standards will have to be developed simultaneously among multiple service suppliers in cloud blockchain design, market creation and customer service support, which promises to facilitate current BCoT-related industries \cite{218}.

Meanwhile, security problems in BCoT can be solved by security improvements in both CoT and blockchain systems. From the CoT point of view, security evaluations and appropriate solutions are vitally important. For example, the work in \cite{223} proposed a trust assessment framework for cloud services named STRAF which takes into account security as a crucial feature to investigate trustworthiness of cloud computing to ensure security of cloud-based IoT applications. Furthermore, a cloud IoT architecture was also presented in \cite{224} in which the trust evaluation mechanism guarantees high security for IoT and enhance system trustworthiness. 
In the perspective of blockchain, there are also some security enhancements. For instance, a mining pool system called SmartPool \cite{225} was proposed to improve transaction verification in blockchain mining to mitigate security bottlenecks, such as 51\% vulnerability, ensuring that the ledger cannot be hacked by increasingly sophisticated attackers. Particularly, recent works \cite{226}, \cite{227} introduced efficient security analysis tools to investigate and prevent threat potentials in order to ensure trustful smart contract execution on blockchain. Such research efforts make contributions to addressing security issues in BCoT environments and improving the overall performance of the system. 

Moreover, many innovative approaches have been considered to enhance privacy for BCoT systems such as encryption methods, trusted cloud computing, efficient user identification, access control, and intention hiding solutions \cite{229}. Recently, an access control architecture was proposed in \cite{230} to improve privacy of IoT data in cloud computing with better data reliability levels inherited by a consensus mechanism. From the blockchain view, anonymity plays a crucial role in ensuring robust privacy for BCoT users. In this regard, user information on blockchain can be hidden efficiently and attackers cannot guess identity of transactions and thus preserve private user information. For example, a recent work in \cite{231} proposes a new solution with the ability to provide anonymity and unlinkability of senders, the privacy of transaction on the blockchain platform. Additionally, the authors in \cite{232} present an anonymous reporting scheme which can ensure the reliability of anonymous reporting information without revealing the identities of IoT users, then preserving the privacy of blockchain systems.

The adoption of expert systems and intelligent tools available on cloud computing may be a good solution to provide intelligence in BCoT-related applications. For example, in BCoT-based smart healthcare, machine learning for smart health assessment systems is useful to support doctors in medical processes \cite{233}, \cite{234}. Meanwhile, in smart cities, big data analytic software available on cloud is very helpful to solve data-related issues such as data collection, processing and visualization for smart services from city environment, citizens and various departments and agencies in the city scale \cite{235}. Specially, a recent research effort \cite{236} was put on the integration of machine learning and blockchain to enable decision making services in a fashion the intelligence of the system is improved while security and reliability are guaranteed.

Some intelligent approaches have been proposed recently to enhance efficiency of resource management in BCoT. For example, the authors in \cite{142} presented a framework for energy-aware resource management in cloud datacentres with blockchain. Machine learning is embedded to smart contracts to optimize energy consumption according to user requests. This solution also has the potentials to achieve significant cost savings in respect with request scheduling and request migration on cloud blockchain. Meanwhile, the issue of resource management for mining blockchain in BCoT is considered in \cite{238} in which resource for computation in the blockchain consensus process is optimized to achieve minimum prices of service usage. It also demonstrates that cloud providers can gain benefits from the optimal resource management with better profits in the proposed public cloud blockchain networks.

\section{\textcolor{black}{Future Directions}}
As BCoT has attracted widespread attention of both academics and industries, its developments are likely to be affected by other technologies. The convergence of BCoT and these technologies can open up a wide range of opportunities to future services and applications. In this section, we will provide insights of such technologies and present the future research directions of integrating BCoT in such technologies to empower both worlds. 
\subsection{{Improving Blockchain Performance for Future BCoT}}
{To realize the full potential of blockchain in future BCoT applications, improving blockchain efficiency is highly important. Due to the security features of blockchain, each blockchain node is often required to verify transactions and authenticate user messages, which may require large computing and storage resources. Blockchain also requires network bandwidth and energy resources from cloud services to implement the mining process. Moreover, current blockchain designs still remain scalability issues from the perspectives of throughput, storage and networking. For example, many current blockchain platforms suffer from a long queueing time for transaction processing due to the block size restriction, which results in high block generation latency \cite{24}. In addition to that, each blockchain node often has to store a copy of complete transaction data which poses a storage burden on the blockchain system. The integration of blockchain into CoT, therefore, will introduce new technical challenges that can negatively affect the overall performance of BCoT systems. Recently, many solutions have been proposed to improve these problems. A research direction is to provide lightweight consensus mechanisms, in order to enhance the blockchain performance by compressing consensus storage \cite{future1} or designing lightweight block validation schemes \cite{future2}. These approaches would help simplify the blockchain mining process to achieve energy savings and latency improvement. Another potential solution is to build off-chain blockchain platforms \cite{future3} by cooperating cloud services with hybrid consensus protocols on top of BCoT networks \cite{future4}. These techniques are capable of mitigating the data processing and storage burden posed on CoT devices and improving the scalability issues of BCoT platforms.  }

\subsection{Machine Learning with BCoT}
\textcolor{black}{The main vision of future BCoT is to provide ubiquitous IoT services with system performance improvements and security enhancements to meet the ever-growing demands of user traffic and emerging services in the future networks, i.e., 5G and beyond. To achieve these goals, strategies for critical BCoT network issues such as networking optimization, system management, resource scheduling are vitally necessary. However, most current solutions have been based on traditional optimization algorithms or centralized designs, which remains some critical challenges \cite{nguyen2020wireless}. For example, the explosion of IoT data volumes in future BCoT ecosystems can make traditional data processing techniques inefficient, and the high dynamics of IoT data traffic make computation and service management become challenging. To overcome these challenges, machine learning has emerged as a highly efficient solution for supporting future BCoT. As an important enabling technology for artificial intelligence, machine learning has been successfully applied in many areas, including computer vision, medical diagnosis, and speech recognition.} The revolution of machine learning technology transforms current BCoT services by enabling its ability to learn from data and provide data driven insights, decision support, and predictions for improving network performances. For example, the work in \cite{235} demonstrates the efficiency of machine learning tools in improving intelligence of BCoT applications. Specially, DRL \cite{241} has recently emerged as one of the most attractive machine learning techniques to solve many critical issues in BCoT networks. Indeed, the integration of DRL can overcome security challenges brought by the dynamic data exchanges among blockchain users and information flow over untrusted mobile environments in cloud-blockchain systems. Our recent works verify that DRL-based offloading for consensus mechanisms in mobile blockchain networks can achieve security improvements for BCoT systems \cite{242}. Obviously, the adoption of machine learning provides more perspectives to evaluate, analyse and deal with existing issues in BCoT scenarios, enabling to boost QoS, security and performance of the whole network. 
\subsection{Big Data in BCoT}

\textcolor{black}{With the rapid development of BCoT applications, big data has emerged as a significant data analytic tool for realizing the potential of the knowledge discovery from huge blockchain IoT data. In the future networks, it is expected that BCoT will witness an exponential growth of data traffic, from the volume, velocity, and variety of blockchain data. Big data can support a number of solutions to facilitating BCoT systems, including storage, data cleaning, and analytics \cite{255}.  Furthermore, big data also supports cleaning services which are understood as a pre-processing phase before big data analytics, mainly used to integrate and improve the quality of big data. More specific, the cleaning service includes two main phases: data integration (data fusion or data aggregation) and data quality management which address issues related to low-quality data such as corrupted data detection in the blockchain data, data redundancy reduction in BCoT data collection services (e.g., blockchain-based sensor networks). Then the analytics service includes all the methods and models for data analysis and processing such as data clustering algorithms and MapReduce processing \cite{256}. For example, data clustering has been applied to characterize the usage and performance characteristics of large peer-to-peer consensus-based systems where blockchain datasets (e.g., Bitcoin data) are aggregated, analysed and visualized to discover unanticipated patterns in blockchain networks. }

In return, BCoT also support big data in terms of better data integrity and privacy preservation to make sure that data analytics in big data is secure. In such contexts, blockchain in BCoT appears as the ideal candidate to solve big data-related issues \cite{251}. Indeed, the decentralized management associated with authentication and reliability of blockchain can provide high security guarantees to big data resources. Specially, blockchain can offer transparency and trustworthiness for sharing of big data among service providers and data owners. By eliminating fear of security bottlenecks, BCoT can enable universal data exchange which empowers large-scale BCoT deployments. Recently, some big data models enabled by blockchain are proposed, such as data sharing with smart contracts \cite{252} or data tracing solutions using blockchain transaction \cite{253}. Such preliminary results show that blockchain can bring various advantages in terms of security and performance to big data applications, which will be promising to the development of both worlds.

\subsection{{BCoT in 5G Networks and Beyond}}
{ The next generations of mobile network (5G and beyond) have revolutionized industry and society by providing an unimaginable level of innovation with some key advantages such as high data rate, low network latency, energy savings, reduced operational costs, higher system throughput, and massive device connectivity. However, the development of new technology architectures in 5G wireless networks such as software defined networking (SDN), network functions virtualisation (NFV), network slicing, device-to-device (D2D) communications, and cloud computing has also raised many security issues \cite{167}. For example, SDN remains some security issues such as forged or faked traffic flows, attacks on vulnerabilities in switches, control plane communications, and controllers, and the lack of trust mechanisms between controllers and management applications \cite{d1}. Moreover, how to provide the integrity between the service providers and platforms to avoid data leakage risks in resource sharing among NFV users and servers is still an open issue \cite{d3}.  In such contexts, the blockchain is able to provide viable security solutions. For example, blockchain can build decentralized authentication mechanisms for SDN to implement decentralized access authorization with smart contracts \cite{d2}. By using shared ledgers, blockchain can also build trust among network entities, e.g., SDN controllers and network users, for reliable communications and secure data exchange. In NFV, blockchain can secure the delivery of network functions and ensure the system integrity against data threats, e.g., malicious VM modifications and data attacks \cite{d4}. 
	
To support future IoT applications, 5G also relies on the network slicing concept, which enables multiple tenants to share the same physical hardware. However, the network slicing operation also remains inter-slice security issues. For example, if the communication link is shared between multiple slices, then a malicious user on one slice could impact other slices, e.g., exploiting the resources or compromising data of the target slice \cite{d5}. Blockchain can be exploited to build reliable end-to-end network slices and allow network slide providers to manage their resource \cite{d6}.  When a slice provider receives a request to establish an end-to-end slice, this request is submitted to the blockchain for authentication using smart contracts. In this way, the resource providers can perform resource trading on contracts with sub-slice components, and the information of sub-slice deployment is immutably recorded and stored in the blockchain. In terms of D2D communications in 5G networks, blockchain is able to build trust between D2D users and ensure transparent and reliable data exchange among different users \cite{d9}.  In this blockchain-based D2D scenario, only resourceful devices (e.g., laptops, powerful smartphones) or edge servers participate in the blockchain mining process, while other lightweight D2D devices only join the network for communications without requiring blockchain mining \cite{d8}.
	
Moreover, blockchain can support well 5G services. As an example, blockchain has been used to achieve trust management for 5G mobile vehicular communication due to its decentralized and immutable characteristics \cite{134}. By using blockchain, the 5G-VANET scheme can detect network attacks and prevent data threats from accessing vehicular ecosystems. Besides, it is proven that blockchain can support secure and flexible key management in 5G IoT networks \cite{169} thanks to its ability to reduce computational complexity and enhance communication security. Moreover, the blockchain along with cloud computing can bring great opportunities to 5G network management. For example, the blockchain can be exploited to build reliable end-to-end network slices and allow network slide providers to manage their resources. The work in \cite{157} uses blockchain for dynamic control of vehicle-to-vehicle and vehicle-to-everything communications in vehicular network slices. Meanwhile, the cloud-native architecture with its programmable networking potentially improves 5G network slicing functions. For instance, the study in \cite{162} demonstrates that the cloud-native model can enable life-cycle slice management to create, orchestrate and optimize the performances of network slices in terms of network resources, end-to-end delay, and data throughput. These research findings are expected to pave the way for the next generation of BCoT-5G networks.}
\section{Conclusions}
In this paper, we have presented an extensive and up-to-date review of the integration of two disruptive technologies: blockchain and CoT, referred to as BCoT, which is becoming increasingly important in industrial applications due to its advantages in security, privacy, and service support. This survey is motivated by the lack of a comprehensive literature review on the development of BCoT systems. We have first discussed the recent advances of BCoT along with the integration motivations and the conceptual BCoT integrated architecture. In particular, we have presented a comprehensive survey on the use of BCoT models in various applied scenarios, ranging from smart healthcare, smart city, smart transportation to industry applications and cloud services. We have further surveyed the emerging BCoT platforms and services that would be useful to application developers and researchers. From the extensive literature review on BCoT applications, we have highlighted some key technical challenges and pointed out possible future directions in BCoT research. 
\bibliography{Ref}
\bibliographystyle{IEEEtran}
\balance
\begin{IEEEbiography}[{\includegraphics[width=1in,height=1.25in,clip,keepaspectratio]{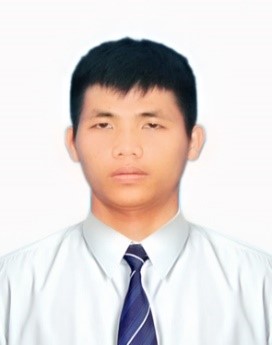}}]{Dinh C. Nguyen}
	(Graduate Student Member, IEEE) is currently pursuing the Ph.D. degree at the School of Engineering, Deakin University, Victoria, Australia. He is also affiliated with the Information Security and Privacy Research Group, CSIRO Data61, Docklands, Melbourne, Australia. His research interests focus on blockchain, deep reinforcement learning, mobile edge/cloud computing, Internet of Things, network security and privacy. He is currently working on blockchain and reinforcement learning for Internet of Things and 5G networks. He has been a recipient of the prestigious Data61 PhD scholarship, CSIRO, Australia.
\end{IEEEbiography}

\begin{IEEEbiography}[{\includegraphics[width=1in,height=1.25in,clip,keepaspectratio]{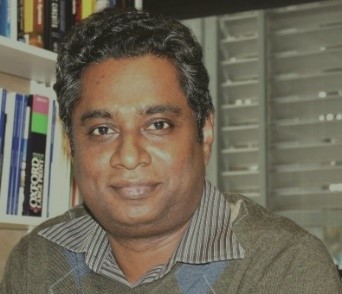}}]{Pubudu N. Pathirana}
	(Senior Member, IEEE) was born in 1970 in Matara, Sri Lanka, and was educated at Royal College Colombo. He received the B.E. degree (first class honors) in electrical engineering and the B.Sc. degree in mathematics in 1996, and the Ph.D. degree in electrical engineering in 2000 from the University of Western Australia, all sponsored by the government of Australia on EMSS and IPRS scholarships, respectively. He was a Postdoctoral Research Fellow at Oxford University, Oxford, a Research Fellow at the School of Electrical Engineering and Telecommunications, University of New South Wales, Sydney, Australia, and a Consultant to the Defence Science and Technology Organization (DSTO), Australia, in 2002. He was a visiting professor at Yale University in 2009. Currently, he is a full Professor and the Director of Networked Sensing and Control group at the School of Engineering, Deakin University, Geelong, Australia and his current research interests include Bio-Medical assistive device design, human motion capture, mobile/wireless networks, rehabilitation robotics and radar array signal processing.
\end{IEEEbiography}

\begin{IEEEbiography}[{\includegraphics[width=1in,height=1.25in,clip,keepaspectratio]{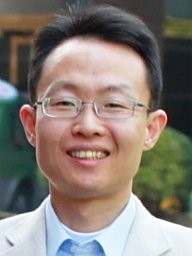}}]{Ming Ding}
	(Senior Member, IEEE) received the B.S. and M.S. degrees (Hons.) in electronics engineering and the Ph.D. degree in signal and information processing from Shanghai Jiao Tong University (SJTU), Shanghai, China, in 2004, 2007, and 2011, respectively. From April 2007 to September 2014, he worked as a Researcher/Senior Researcher/Principal Researcher at the Sharp Laboratories of China, Shanghai. He also served as the Algorithm Design Director and the Programming Director for a system-level simulator of future telecommunication networks in Sharp Laboratories of China for more than seven years. He is currently a Senior Research Scientist with the CSIRO Data61, Sydney, NSW, Australia. His research interests include information technology, data privacy and security, machine learning and AI. He has authored over 100 articles in IEEE journals and conferences, all in recognized venues, and around 20 3GPP standardization contributions, and a Springer book Multi-Point Cooperative Communication Systems: Theory and Applications. He holds 21 U.S. patents and co-invented another more than 100 patents on 4G/5G technologies in CN, JP, KR, EU. He is an Editor of the IEEE TRANSACTIONS ON WIRELESS COMMUNICATIONS and the IEEE Wireless Communications Letters. Besides, he is or has been a Guest Editor/CoChair/Co-Tutor/TPC Member of several IEEE top-tier journals/conferences, such as the IEEE JOURNAL ON SELECTED AREAS IN COMMUNICATIONS, IEEE Communications Magazine, and the IEEE GLOBECOM Workshops. 
\end{IEEEbiography}

\begin{IEEEbiography}[{\includegraphics[width=1in,height=1.25in,clip,keepaspectratio]{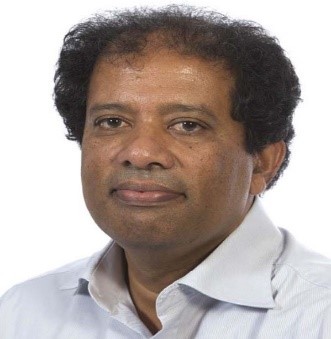}}]{Aruna Seneviratne}
	(Senior Member, IEEE) is currently a Foundation Professor of telecommunications with the University of New South Wales, Australia, where he holds the Mahanakorn Chair of telecommunications. He has also worked at a number of other Universities in Australia, U.K., and France, and industrial organizations, including Muirhead, Standard Telecommunication Labs, Avaya Labs, and Telecom Australia (Telstra). In addition, he has held visiting appointments at INRIA, France. His current research interests are in physical analytics: technologies that enable applications to interact intelligently and securely with their environment in real time. Most recently, his team has been working on using these technologies in behavioral biometrics, optimizing the performance of wearables, and the IoT system verification. He has been awarded a number of fellowships, including one at British Telecom and one at Telecom Australia Research Labs.
\end{IEEEbiography}
\end{document}